%% file: main.tex
\documentclass[prx, twocolumn, superscriptaddress, dvipsnames]{revtex4-2}
\pdfoutput=1
\usepackage[utf8]{inputenc}
\usepackage[english]{babel}
\usepackage{amsmath}
\usepackage{amssymb}
\usepackage{amsthm}
\usepackage{amsfonts}
\usepackage{graphicx}
\usepackage{bbm}
\usepackage{makecell}
\usepackage{xcolor}
\usepackage{bm}
\usepackage{here}
\usepackage{soul}
\usepackage{braket}
\usepackage{bbold}

\usepackage[colorlinks,allcolors=blue]{hyperref}
\usepackage[noabbrev, capitalize]{cleveref}

\newtheorem{thm}{Theorem}[section]

\newtheorem{prop}[thm]{Proposition}

\creflabelformat{equation}{#2\textup{#1}#3}
\theoremstyle{definition}
\newtheorem{defn}[thm]{Definition}

\usepackage{tikz}
\usetikzlibrary{decorations.pathreplacing,calligraphy,decorations.markings}

\definecolor{whampdocolor}{HTML}{FCDE70}
\definecolor{whampdocolorw}{HTML}{EEF7FF}
\definecolor{whitetensorcolor}{HTML}{F8F8F8}
\definecolor{lcolor}{HTML}{D9EAFD}

\input{my_tikz}

\usetikzlibrary{decorations.pathmorphing, decorations.markings, arrows, arrows.meta, shapes, patterns}
\tikzset{baseline={([yshift=-.5ex]current bounding box.center)}}
\tikzset{every path/.style={ line width=0.5pt, line cap=round }}
\colorlet{Virtual}{RedOrange}
\tikzstyle{bevel} = [ preaction = { draw, white, line width=3pt,  line cap = round } ]
\tikzstyle{bevel wide} = [ preaction = { draw, white, line width=4pt,  line cap = round } ]
\tikzstyle{symb} = [ draw=black, fill=black, line width=0.4pt, inner sep=1.5pt ]
\tikzstyle{mysymb} = [ draw=black, fill=white, circle, line width=0.3pt, inner sep=1pt, font=\small ] 
\tikzstyle{symb large} = [ inner sep=2.1pt ]
\tikzstyle{symb small} = [ inner sep=1pt   ]
\tikzstyle{symb tiny} = [ inner sep=0.8pt ]
\tikzstyle{symb fdisk} = [ circle ]
\tikzstyle{symb disk} = [ circle ]
\tikzstyle{symb square} = [ rectangle ]
\tikzstyle{symb fsquare} = [ rectangle ]
\tikzstyle{Msymb}=[draw=black, fill=whampdocolor, circle, inner sep=1pt, font=\small]
\tikzstyle{Nsymb}=[draw=black, fill=whampdocolorw, circle, inner sep=1pt, font=\small]
\tikzstyle{-mid} = [ decoration={ markings, mark = at position 0.50*\pgfdecoratedpathlength+0.6*3pt with \arrow{>[width=2pt]} }, postaction={decorate} ]
\tikzstyle{mid-} = [ decoration={ markings, mark = at position 0.50*\pgfdecoratedpathlength+0.6*3pt with \arrow{<[width=2pt]} }, postaction={decorate} ]

\newcommand\subsetsim{\mathrel{%
  \ooalign{\raise0.2ex\hbox{$\subset$}\cr\hidewidth\raise-0.8ex\hbox{\scalebox{0.9}{$\sim$}}\hidewidth\cr}}}

\newcommand{\ra}{\rightarrow}

\newcommand{\tr}{\mathrm{Tr}}
\newcommand{\bo}{\mathbbm{1}}

\newcommand{\rd}{\mathrm{d}}

\newcommand{\mA}{\mathcal{A}}
\newcommand{\mB}{\mathcal{B}}
\newcommand{\dg}{\dagger}

\newcommand{\tc}{\mathrm{TC}}
\newcommand{\ds}{\mathrm{DS}}
\newcommand{\tE}{\tilde{\mathcal{E}}}
\newcommand{\tR}{\tilde{\mathcal{R}}}
\newcommand{\N}{\mathsf{N}}
\newcommand{\drangle}{\rangle\!\rangle}
\newcommand{\Ainn}{\mathsf{A}}

\definecolor{yuhan}{rgb}{0.9, 0, 0.5}

\allowdisplaybreaks

\begin{document}
\title{Establishing Mixed-State Phase Equivalence beyond \\ Renormalization Fixed Points}


\author{Yuhan Liu}
\affiliation{Max Planck Institute of Quantum Optics, Hans-Kopfermann-Str. 1, Garching 85748, Germany}
\affiliation{Munich Center for Quantum Science and Technology (MCQST), Schellingstr. 4, 80799 M{\"{u}}nchen, Germany}


\begin{abstract}
Understanding mixed-state quantum phases is a central challenge in the era of quantum simulation, where many existing studies focus on renormalization fixed points. In this work, we move beyond the renormalization fixed-point paradigm by constructing a quantum phase transition connecting two distinct one-dimensional fixed points, both exhibiting finite conditional mutual information and one of which is intrinsically nontrivial. We analytically establish phase equivalence within each of the two phases by explicitly constructing low-depth, quasi-local channel circuits that connect states within each phase. Crucially, our approach leverages the parent Lindbladian construction to generate the desired channel circuits. We further demonstrate that this framework generalizes naturally to a broad class of intrinsically nontrivial mixed-state quantum phases. Our method establishes a framework for rigorously analyzing phase equivalence of intrinsically non-trivial mixed states beyond the renormalization fixed points. 
\end{abstract}

\maketitle

\section{Introduction}

Understanding quantum phases and phase transitions is a central problem in physics. Two pure states are said to be in the same phase if one can be efficiently transformed into the other by a low-depth, quasi-local unitary circuit~\cite{chen2010local}. For two pure states in distinct phases, one may construct a continuous path connecting them that passes through a phase transition~\cite{schuch2011classifying,wolf2006quantum,jones2021skeletons}. This path is physical in the sense that each state on the path can be realized as the ground state of a local Hamiltonian. Along the path, the spectral gap remains open away from the transition point and closes at the transition.


Mixed-state quantum phases have attracted significant recent interest, driven by advances in quantum simulation. Analogously, two mixed states are said to belong to the same phase if one can be transformed into the other via a low-depth, quasi-local circuit of quantum channels~\cite{Coser2019classificationof,sang2024mixed,ruiz2024matrix}. Remarkably, intrinsically nontrivial mixed-state quantum phases exist even in one-dimensional systems~\cite{lessa2025mixed,ruiz2024matrix,sun2025anomalous}, where no pure-state counterparts arise. This class of intrinsically nontrivial mixed states exhibits an anomalous nonlocal symmetry, where the symmetry admits a matrix product operator (MPO) representation. Despite the presence of the symmetry, these states remain intrinsically nontrivial and are not classified as symmetry-protected phases. Beyond one dimension, nontrivial mixed-state phases have also been identified in higher-dimensional systems~\cite{ellison2025toward,sohal2025noisy}, as well as in settings with symmetry protection~\cite{de2022symmetry,ma2023average,guo2025locally}. 

Much of the efforts on intrinsically non-trivial mixed states focus on the renormalization fixed points~\cite{cirac2017matrix}, which have zero correlation length and rich analytical structures~\cite{molnar2022matrix,ruiz2024matrix,liu2026parent,liu2025trading}. Going beyond the renormalization fixed points, it is natural to consider two mixed states in distinct phases and construct a continuous path connecting them that passes a phase transition. In earlier works, mixed-state phases and transitions are characterized through non-analytical changes of local observables~\cite{onsager1944crystal,diehl2008quantum,diehl2010dynamical}. Beyond this paradigm, an important example arises in quantum memories (for example the toric code) under local decoherence. In the context of quantum error correction, this transition corresponds to a decoding threshold~\cite{dennis2002topological,wang2003confinement}, beyond which the success probability of any decoding algorithm vanishes. Such phase distinctions can be diagnosed using entanglement measures~\cite{fan2024diagnostics}. The entanglement probes have also proven effective in other models~\cite{lu2020detecting,lu2025holographic}. Besides, mixed-state phase transition is also studied using anyon condensation in symmetry topological field theory~\cite{luo2025topological}.

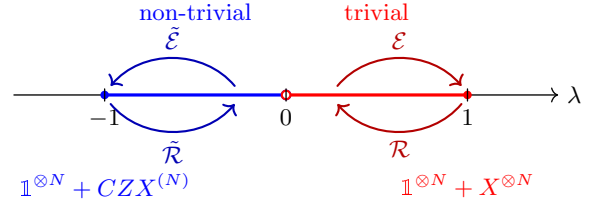
\begin{figure}[t]
  \centering
  \begin{tikzpicture}[scale=0.8]
  \def\off{0.15};

    \draw[->] (-4.5,0) -- (4.5,0) node[right] {$\lambda$};

    \draw[very thick, blue] (-3,0) -- (0,0);
    \fill[blue] (-3,0) circle (2pt);          
    \draw[blue, fill=white, thick] (0,0) circle (2pt); 

    \draw[very thick, red] (0,0) -- (3,0);
    \draw[red, fill=white, thick] (0,0) circle (2pt); 
    \fill[red] (3,0) circle (2pt);             

    \draw (-3., 0.1) -- (-3.,-0.1) node[below] {$-1$};
    \draw (0, 0.1) -- (0,-0.1)   node[below] {$0$};
    \draw (3, 0.1) -- (3,-0.1)   node[below] {$1$};

    \node[blue] at (-3.,-1.5) {$\bo^{\otimes N}+CZX^{(N)}$};
    \node[red] at (3.,-1.5) {$\bo^{\otimes N}+X^{\otimes N}$};

    \node[above=8pt, blue]  at (-1.5, 0.75) {non-trivial};
    \node[above=8pt, red]   at ( 1.5, 0.75) {trivial};

    \draw[->, thick, blue!70!black] (-0.85, \off) to[bend right=50] node[above] {$\tilde{\mathcal{E}}$} (-2.9, \off);

    \draw[->, thick, blue!70!black] (-2.9, -\off) to[bend right=50] node[below] {$\tilde{\mathcal{R}}$} (-0.85, -\off);

    \draw[->, thick, red!70!black] (0.85, \off) to[bend left=50] node[above] {$\mathcal{E}$} (2.9, \off);

    \draw[->, thick, red!70!black] (2.9, -\off) to[bend left=50] node[below] {$\mathcal{R}$} (0.85, -\off);

  \end{tikzpicture}
  \caption{Mixed-state phase diagram as a function of $\lambda$. The intrinsically non-trivial phase corresponds to $-1 \leq \lambda < 0$ and the trivial phase to $0 < \lambda \leq 1$, with $\lambda=\mp 1$ being the mixed-state renormalization fixed points, $\rho(-1)\propto \bo^{\otimes N}+CZX^{(N)}$ and $\rho(1)\propto \bo^{\otimes N}+X^{\otimes N}$. The phase equivalence within each phase is proven by constructing the four channel circuits $\mathcal{E}$ and $\mathcal{R}$, $\bar{\mathcal{E}}$ and $\bar{\mathcal{R}}$.  }
  \label{fig:phase-diagram}
\end{figure}

For more general mixed-state systems, establishing phase equivalence and identifying phase transitions remain challenging, as the corresponding phases may not be distinguishable using either local observables or entanglement measures.
To prove the phase equivalence by definition, one needs to show the existence of a valid channel circuit. An important progress is by noticing the relation between recoverability and conditional mutual information (CMI)~\cite{sang2025stability,junge2018universal}. In particular, when a family of mixed states has decaying CMI, the recovery channel can be constructed explicitly. Nevertheless, for a class of mixed states~\cite{lessa2025mixed,ruiz2024matrix,sun2025anomalous}, CMI is finite even at the renormalization fixed point. 
Moving beyond the renormalization fixed points for this class of states and proving the phase equivalence is thus an open question.

In this paper, we address this question by constructing a quantum phase transition between mixed-state renormalization fixed points, both characterized by a finite CMI and one of which is intrinsically nontrivial. Such a path of mixed states is physical in the sense that each state along the path can be realized as the fixed point of a completely-positive map~\cite{cirac2011entanglement,schuch2013topological,cirac2017matrix,lu2025holographic}. We establish phase equivalence (see Fig.~\ref{fig:phase-diagram}) by extending the method of~\cite{sang2025stability}  and explicitly constructing the required channel circuits. The key insight underlying our proof is that the parent Lindbladian~\cite{liu2026parent} provides a canonical protocol for generating the desired channel circuits that map non–fixed-point states to fixed-point states. Importantly, these channel circuits preserve the MPO symmetry of the mixed state by construction, thereby enabling the explicit construction of recovery channels that map fixed-point states back to non–fixed-point states. We illustrate our approach using a concrete model with anomalous $\mathbb{Z}_2$ MPO symmetry, and then extend to the general setting. Our method establishes a framework for rigorously determining phase equivalence of intrinsically non-trivial mixed states beyond the renormalization fixed points.

\section{Preliminary}
In this section, we introduce the necessary preliminaries. We begin by defining phase equivalence for mixed states, and then introduce the conditional mutual information (CMI), an entanglement measure that plays a central role in characterizing phase equivalence. We next introduce matrix product operators (MPOs), which provide a powerful framework for describing mixed states and their associated symmetries.

\subsection{Definition of phase equivalence}
Given a one-dimensional lattice $\mathcal{L}=\{1,2,\cdots,N\}$ and the Hilbert space $\mathcal{H}=\bigotimes_{i\in{\mathcal{L}}} \mathbb{C}^2$, we consider the density matrices $\rho$ supported on $\mathcal{H}$. 

A channel circuit is a completely positive trace-preserving (CPTP) map of the form
\begin{equation}
\mathcal{E} = \prod_{l=1}^{D_{\mathcal{E}}} \ \prod_{x} \mathcal{E}_{l,x},
\end{equation}
where $l$ labels the circuit layers, $x$ labels the channels within each layer, and $D_{\mathcal{E}}$ is the number of layers. We require that, for each fixed layer $l$, the supports of the channels $\{\mathcal{E}_{l,x}\}_x$ are pairwise disjoint, and each $\mathcal{E}_{l,x}$ is supported on a simply connected region. We denote the combined index $(l,x)$ as $I$. 

Let $r_{\mathcal{E}} := \max_{l,x} |\mathrm{supp}(\mathcal{E}_{l,x})|$ denote the maximal support size. Given a channel circuit $\mathcal{E}$ that is well-defined for all system sizes $N$. The circuit is called \emph{strictly local} if $r_{\mathcal{E}} = O(1)$ (independent of the system size $N$), and \emph{quasi-local} if $r_{\mathcal{E}} = \mathrm{poly}\log(N)$. The circuit is called \textit{finite-depth} if $D_{\mathcal{E}}=O(1)$ and \textit{low-depth} if $D_{\mathcal{E}}=\mathrm{poly}\log(N)$.

\begin{defn}[Mixed-state phase equivalence]
\label{def:phase-equiv}
Given two states $\rho_0$ and $\rho_1$ that are well-defined for all system sizes $N$.
    The state $\rho_0$ can be brought fast to another state $\rho_1$, and we write $\rho_0\ra \rho_1$, if there exists a low-depth quasi-local channel circuit $\mathcal{E}$, such that 
    \begin{equation}
        \| \mathcal{E}(\rho_0)-\rho_1\|_1\leq \epsilon
    \end{equation}
    with $\epsilon\ra 0$ in the thermodynamic limit $N\ra \infty$ (the norm using above is Schatten 1-norm). 

    Two states belong to the same phase if there exist two low-depth quasi-local channel circuits $\mathcal{E},\mathcal{R}$ such that $\rho_0\ra \rho_1$ via $\mathcal{E}$ and $\rho_1 \ra \rho_0$ via $\mathcal{R}$. 
\end{defn}

In particular, if $\rho$ is in the same phase as the product state, we say $\rho$ is in the trivial phase. 
We note that we adopt the discrete evolution definition of mixed-state phase equivalence~\cite{ruiz2024matrix,sang2024mixed} instead of the continuous evolution definition using the Lindbladian~\cite{Coser2019classificationof}. The distinction does not affect the results in this paper, and the discrete version is used for convenience. 

We comment that this definition admits a natural physical interpretation in terms of causal structure. A finite-depth strictly-local circuit generates a strict light cone~\cite{chen2010local}, and for quasi-local, low-depth circuits the light cone grows at most poly-logarithmically with system size, which cannot generate long-range correlations. This behavior is the discrete analogue of the Lieb–Robinson bound, which establishes an emergent finite velocity for information propagation under local Hamiltonian~\cite{Bravyi2006Lieb-Robinson} or Lindbladian evolution~\cite{poulin2010LiebRobinson}.

\subsection{Conditional mutual information and reversability}
Given a state $\rho$ supported on $\mathcal{H}$, its reduced density matrix on region $A\subseteq \mathcal{L}$ is $\rho_A:=\tr_{\bar{A}}\rho$ where $\bar{A}=\mathcal{L}\setminus A$ denotes the complement of $A$. The von Neumann entanglement entropy is $S(A):=-\tr(\rho_A \log\rho_A)$ and the conditional mutual information (CMI) is defined as $I(A:C|B)=S(AB)+S(BC)-S(B)-S(ABC)$.  

Physically, the CMI quantifies the amount of correlation between regions $A$ and $C$ that is not mediated by $B$. A small CMI indicates that $B$ approximately screens the correlations between $A$ and $C$, so that any remaining correlations are weak and can be recovered from $B$ up to a small error~\cite{fawzi2015quantum,sutter2016universal}.

\begin{defn}[Finite Markov length]
    Let $\rho$ be a density matrix supported on $\mathcal{H}$, and define $A=\{l_A,\cdots,r_A\},B=\{l_A-d,\cdots,l_A-1\}\cup\{r_A+1,\cdots,r_A+d\}$, and $C=\mathcal{L}\setminus (A\cup B)$, i.e., two simply connected regions A and C are separated by region $B$ of width $d$. We say $\rho$ has Markov length $\xi$ if its CMI satisfies
    \begin{equation}
        I(A:C|B)\leq \mathrm{poly}(|A|,|C|)e^{-d/\xi}.
    \end{equation}
    Let $\rho$ be well-defined for all system sizes $N$. We say $\rho$ has $\xi$-finite Markov length if the Markov length for any $N$ is upper bounded by $\xi$~\cite{sang2025stability}.
\end{defn}

The behavior of conditional mutual information imposes strong constraints on the structure of a density matrix $\rho$. For instance, when the CMI vanishes exactly, it implies a nontrivial Hilbert space decomposition~\cite{hayden2004structure} and $\rho$ is a Markov chain. More generally, a decaying CMI underlies the approximation of density matrices as thermal states~\cite{leifer2008quantum,brown2012quantum,kato2019quantum}. In the context of channel circuits, a decaying CMI guarantees approximate recoverability, leading to the following result~\cite{sang2025stability,junge2018universal}:

\begin{prop}[Recoverability]
\label{prop:reverse}
    Let $\rho$ be well-defined for all system sizes $N$ and with $\xi$-finite Markov length. Let $\mathcal{E}=\prod_{I\in S}\mathcal{E}_I$ be a circuit of strictly-local finite-depth quantum channel, such that the Markov length of $\prod_{I\in S'}\mathcal{E}_I(\rho)$ of any $S'\subseteq S$ is upper bounded by $\xi$. Then, there exists a circuit of quasi-local low-depth quantum channel $\mathcal{R}=\prod_{J}\mathcal{R}_J$, where $r=\max_{J} |\mathrm{supp}(\mathcal{R}_{J})|$ is of order
    \begin{equation}
        r\sim \xi \log\left(
        \frac{\mathrm{poly}(N)}{\epsilon}
        \right) 
    \end{equation}
    such that
    \begin{equation}
        \|\mathcal{R}\circ\mathcal{E}(\rho) - \rho \|_1\leq \epsilon.
    \end{equation}
\end{prop}

In particular, choosing $\epsilon\sim 1/N$ leads to that $\rho$ and $\mathcal{E}(\rho)$ are in the same phase under~\cref{def:phase-equiv}. 

\subsection{Matrix Product Operator}
Matrix product operator is a very useful tool in analyzing phases of matter, as we introduce below~\cite{perez2006matrix,schollwock2011density,cirac2021matrix}. 

Consider a rank-4 tensor $A$ with graphical representation, 
\begin{align}
    \left(A^{ij}\right)_{\alpha\beta} =
        \begin{array}{c}
        \begin{tikzpicture}[scale=1.,baseline={([yshift=-0.65ex] current bounding box.center)}]
		\draw (-0.75,0) node {$\alpha$};
		\draw (0.75,0) node {$\beta$};
		\draw (0,0.75) node {$i$};
        \draw (0,-0.75) node {$j$};
        \mposite{(0,0)}{2};
        \end{tikzpicture}
        \end{array}
        \in \mathbb C
\end{align}
where $i,j=1,2,\cdots,d$ are physical space indices and $\alpha,\beta=1,2,\cdots,D$ are auxiliary space indices. Given a tensor $A$, one can concatenate $N$ copies to form a uniform matrix product operator (MPO), 
\begin{equation}
\begin{aligned}
     O^{(N)}(A)=&\sum_{\lbrace i,j\rbrace}\tr\left( A^{i_1 j_1} A^{i_2 j_2}\cdots A^{i_N j_N}\right)\\
     &\quad  |i_1 i_2 \cdots i_N\rangle\langle j_1 j_2 \cdots j_N|,
\end{aligned}
\end{equation}
or graphically,
\begin{equation}
\label{eqn:graph-O}
   O^{(N)}(A)=
    \begin{array}{c}
\begin{tikzpicture}[scale=1]
    \pgfmathsetmacro{\x}{0.494}       
    \pgfmathsetmacro{\dx}{0.988}      
    \pgfmathsetmacro{\xlast}{2.964}   
    \pgfmathsetmacro{\vl}{0.247}      
    \pgfmathsetmacro{\yo}{0.564}      
    \pgfmathsetmacro{\ybot}{0.318}    
    \pgfmathsetmacro{\ytop}{1.129}    
    \pgfmathsetmacro{\yleg}{0.811}    
    \pgfmathsetmacro{\ybez}{0.071}    

    \draw[Virtual] (\x+\vl, \yo)    -- (\x+\dx-\vl, \yo);   
    \draw[Virtual] (\x+\dx+\vl, \yo)-- (\x+\dx+\vl+\vl, \yo); 
    \node[anchor=center] at (\x+\dx+2*\vl+0.249, \yo+0.097) {$\overset{N}\cdots$};
    \draw[Virtual] (\xlast-\vl, \yo) -- (\xlast-\vl-\vl, \yo); 
    \draw[Virtual] (\x, \yo) -- (\x-\vl, \yo);               

    \draw[Virtual] (\x-\vl, \yo) arc[start angle=87.734, end angle=270, radius=\vl]
        -- (\xlast+\vl, \ybez) arc[start angle=-90, end angle=90, radius=\vl];

    \draw[-mid] (\x,      \yleg) -- (\x,      \ytop);
    \draw[-mid] (\x+\dx,  \yleg) -- (\x+\dx,  \ytop);
    \draw[-mid] (\xlast,  \yleg) -- (\xlast,  \ytop);

    \draw[Virtual] (\xlast+\vl, \yo) -- (\xlast, \yo);

    \draw[bevel, -mid] (\x,     0) -- (\x,     \ybot);
    \draw[bevel, -mid] (\x+\dx, 0) -- (\x+\dx, \ybot);
    \draw[bevel, -mid] (\xlast, 0) -- (\xlast, \ybot);

    \Osymb{(\x,     \yo)};
    \Osymb{(\x+\dx, \yo)};
    \Osymb{(\xlast, \yo)};
\end{tikzpicture}
        \end{array}
        \,.
\end{equation}

MPOs find wide applications in the study of quantum phases. For instance, they provide an efficient representation of local Hamiltonians~\cite{schollwock2011density}. When supplemented with positivity constraints, they describe matrix product density operators~\cite{verstraete2004matrix,zwolak2004mixed}. Moreover, MPOs offer a natural framework for representing non-local symmetries, including anomalous~\cite{seifnashri2025disentangling,tu2026anomalies,shirley2025anomaly} and non-invertible symmetries. In fact, the presence of MPO symmetries is a characteristic feature of two-dimensional topologically ordered systems~\cite{csahinouglu2021characterizing,BULTINCK2017anyons} as well as symmetry-protected topological phases~\cite{Williamson2016matrix}. 

In our context, the intrinsically nontrivial mixed state $\rho$ admits an MPO symmetry $O$, satisfying
\begin{equation}
O\rho = \rho,
\end{equation}
where $O$ is anomalous. The presence of this MPO symmetry can be understood from a bulk-boundary correspondence~\cite{cirac2011entanglement,schuch2013topological}: $\rho$ can be realized as the boundary state of a two-dimensional system with topological order that is described by a projected entangled-pair state (PEPS), a class of tensor network states (see \cref{app:explicit} for details). Therefore, the MPO symmetry of $\rho$ is inherited from the MPO symmetry for the virtual degrees of freedom of the PEPS tensor. 

\section{Model construction}
We now construct a model of quantum phase transition between the following two mixed-state renormalization fixed points,
\begin{equation}
    \rho_{\tc}=\frac{1}{2^N}(\bo^{\otimes N}+X^{\otimes N})
\end{equation}
and
\begin{equation}
    \rho_{\ds}=\frac{1}{2^N}(\bo^{\otimes N}+CZX^{(N)})
\end{equation}
where $CZX^{(N)}:=\prod_{i=1}^N (CZ)_{i,i+1} \prod_{i=1}^N X_i$. These two density matrices have $\mathbb{Z}_2$ MPO symmetries
\begin{equation}
    \begin{aligned}
        X^{\otimes N} \rho_\tc&=\rho_\tc,\\
        CZX^{(N)} \rho_\ds & = \rho_\ds
    \end{aligned}
\end{equation}
respectively. In particular, $CZX^{(N)}$ is an anomalous $\mathbb{Z}_2$ MPO symmetry associated with a non-trivial 3-cocycle of $\mathbb{Z}_2$ group~\cite{chen2011two,seifnashri2025disentangling,tu2026anomalies,shirley2025anomaly} (and $X^{\otimes N}$ can be viewed as an non-anomalous $\mathbb{Z}_2$ MPO symmetry associated with trivial 3-cocycle). One can prove that $\rho_\tc$ is in the same phase as the product state under~\cref{def:phase-equiv}, and therefore in the trivial phase~\cite{ruiz2024matrix,liu2026parent}; while there exists a no-go theorem to connect $\rho_\ds$ to a product state~\cite{lessa2025mixed}, thus $\rho_\ds$ is in the non-trivial phase. Both $\rho_\tc$ and $\rho_\ds$ admit a finite CMI.

To construct a path of mixed states $\rho(\lambda),-1\leq \lambda\leq 1$ such that $\rho(-1)=\rho_{\ds}$ and $\rho(1)=\rho_\tc$, we consider the path of completely-positive (CP) maps $T(\lambda)$ such that $\rho(\lambda)$ is a fixed point of $T(\lambda)$. Specifically, we consider the CP map $T(\lambda)$ that takes the form of a matrix product CP map generated by a single tensor $E(\lambda)$. $E(\lambda)$ has a direct sum decomposition in the virtual bond, 
\begin{equation}
    E(\lambda)^{iji'j'}=E_1(\lambda)^{iji'j'}\oplus E_2(\lambda)^{iji'j'}.
\end{equation}
The first piece is
\begin{equation}
\label{eqn:E1}
    \left[E_1(\lambda)\right]^{iji'j'}_{\alpha\alpha'\beta\beta'}=\begin{array}{c}
        \begin{tikzpicture}[scale=1., baseline={([yshift=-0.65ex] current bounding box.center)}]
    \pgfmathsetmacro{\rleg}{0.35}
    \pgfmathsetmacro{\ext}{0.2}
    \def\rdot{0.03}

    \opleg{(0,0)}{0.22}{$O_D$}{\rleg};

    \coordinate (TR) at ({\rleg*cos(45)},  {\rleg*sin(45)});   
    \coordinate (TL) at ({\rleg*cos(135)}, {\rleg*sin(135)});  
    \coordinate (BR) at ({\rleg*cos(-45)}, {\rleg*sin(-45)});  
    \coordinate (BL) at ({\rleg*cos(225)}, {\rleg*sin(225)});  

   \coordinate (JTR) at ({\rleg*cos(45)+1.5*\ext},  {\rleg*sin(45)+\ext*0.707});
    \coordinate (JTL) at ({\rleg*cos(135)-1.5*\ext}, {\rleg*sin(135)+\ext*0.707});
    \coordinate (JBR) at ({\rleg*cos(-45)+1.5*\ext}, {\rleg*sin(-45)-\ext*0.707});
    \coordinate (JBL) at ({\rleg*cos(225)-1.5*\ext}, {\rleg*sin(225)-\ext*0.707});

    \draw[\mthick, Virtual] (TR) to[out=45,  in=180] (JTR);
    \draw[\mthick, Virtual] (JTR) -- ++({\ext},0) node[right] {\scriptsize $\beta$};                          
    \draw[\mthick, dashed]          (JTR) to[out=45,  in=270] ++({ \ext*0.707}, {2.*\ext}) node[above] {\scriptsize $j$}; 
    \draw[\mthick]  (JTR) to[out=135, in=270] ++({-\ext*0.707}, {2.*\ext}) node[above] {\scriptsize $i$}; 
    \filldraw (JTR) circle[radius=\rdot];

    \draw[\mthick, Virtual] (TL) to[out=135, in=0]   (JTL) node[left] {\scriptsize $\alpha$};

    \draw[\mthick, Virtual] (BR) to[out=-45, in=180] (JBR);
    \draw[\mthick, Virtual] (JBR) -- ++({\ext},0) node[right] {\scriptsize $\beta'$};                           
    \draw[\mthick, dashed]          (JBR) to[out=-45, in=90] ++({ \ext*0.707},{-2.*\ext}) node[below] {\scriptsize $j'$}; 
    \draw[\mthick]  (JBR) to[out=-135, in=90] ++({-\ext*0.707},{-2.*\ext}) node[below] {\scriptsize $i'$}; 
    \filldraw (JBR) circle[radius=\rdot];

    \draw[\mthick, Virtual] (BL) to[out=225, in=0]   (JBL) node[left] {\scriptsize $\alpha'$};
\end{tikzpicture}
    \end{array}
\end{equation}
where the black dots represent delta functions $\delta_{ijkl}=1$ if $i=j=k=l$ and  $\delta_{ijkl}=0$ otherwise; and the components of $O_D(\lambda)$ are (the order of indices is $\alpha\beta\alpha'\beta'$)
\begin{equation}
\begin{aligned}
    &(O_D)_{0011}=(O_D)_{1100}=(O_D)_{0101}=(O_D)_{1010}=\lambda^2\\
    &(O_D)_{\alpha\beta\alpha'\beta'}=1\quad\text{other components}. 
\end{aligned}
\end{equation}
The second piece $E_2$ is defined by acting an MPO tensor on $E_1$, where for $0\leq \lambda\leq 1$
\begin{equation}
    E_2(\lambda)=\begin{array}{c}
        \begin{tikzpicture}[scale=1., baseline={([yshift=-0.65ex] current bounding box.center)}]
    \pgfmathsetmacro{\rleg}{0.35}
    \pgfmathsetmacro{\ext}{0.2}
    \def\rdot{0.03}

    \opleg{(0,0)}{0.22}{$O_D$}{\rleg};

    \coordinate (TR) at ({\rleg*cos(45)},  {\rleg*sin(45)});   
    \coordinate (TL) at ({\rleg*cos(135)}, {\rleg*sin(135)});  
    \coordinate (BR) at ({\rleg*cos(-45)}, {\rleg*sin(-45)});  
    \coordinate (BL) at ({\rleg*cos(225)}, {\rleg*sin(225)});  

   \coordinate (JTR) at ({\rleg*cos(45)+1.5*\ext},  {\rleg*sin(45)+\ext*0.707});
    \coordinate (JTL) at ({\rleg*cos(135)-1.5*\ext}, {\rleg*sin(135)+\ext*0.707});
    \coordinate (JBR) at ({\rleg*cos(-45)+1.5*\ext}, {\rleg*sin(-45)-\ext*0.707});
    \coordinate (JBL) at ({\rleg*cos(225)-1.5*\ext}, {\rleg*sin(225)-\ext*0.707});

    \draw[\mthick, Virtual] (TR) to[out=45,  in=180] (JTR);
    \draw[\mthick, Virtual] (JTR) -- ++({\ext},0);                          
    \draw[\mthick, dashed]          (JTR) to[out=45,  in=270] ++({ \ext*0.707}, {4.*\ext}); 
    \draw[\mthick]  (JTR) to[out=135, in=270] ++({-\ext*0.707}, {4.*\ext}) ; 
    \filldraw (JTR) circle[radius=\rdot];
     \op{{\rleg*cos(45)+1.5*\ext-0.707*\ext},  {\rleg*sin(45)+3*\ext}}{0.15}{\scriptsize $X$};

    \draw[\mthick, Virtual] (TL) to[out=135, in=0]   (JTL) ;

    \draw[\mthick, Virtual] (BR) to[out=-45, in=180] (JBR);
    \draw[\mthick, Virtual] (JBR) -- ++({\ext},0);                           
    \draw[\mthick, dashed]          (JBR) to[out=-45, in=90] ++({ \ext*0.707},{-4.*\ext}); 
    \draw[\mthick]  (JBR) to[out=-135, in=90] ++({-\ext*0.707},{-4.*\ext}); 
    \filldraw (JBR) circle[radius=\rdot];

    \filldraw (JTR) circle[radius=\rdot];
     \op{{\rleg*cos(45)+1.5*\ext-0.707*\ext},  {-\rleg*sin(45)-3*\ext}}{0.15}{\scriptsize $X$};

    \draw[\mthick, Virtual] (BL) to[out=225, in=0]   (JBL) ;
\end{tikzpicture}
    \end{array}
\end{equation}
and for $-1\leq \lambda\leq 0$
\begin{equation}
    E_2(\lambda)=\begin{array}{c}
        \begin{tikzpicture}[scale=1., baseline={([yshift=-0.65ex] current bounding box.center)}]
    \pgfmathsetmacro{\rleg}{0.35}
    \pgfmathsetmacro{\ext}{0.2}
    \def\rdot{0.03}

    \opleg{(0,0)}{0.22}{$O_D$}{\rleg};

    \coordinate (TR) at ({\rleg*cos(45)},  {\rleg*sin(45)});   
    \coordinate (TL) at ({\rleg*cos(135)}, {\rleg*sin(135)});  
    \coordinate (BR) at ({\rleg*cos(-45)}, {\rleg*sin(-45)});  
    \coordinate (BL) at ({\rleg*cos(225)}, {\rleg*sin(225)});  

   \coordinate (JTR) at ({\rleg*cos(45)+1.5*\ext},  {\rleg*sin(45)+\ext*0.707});
    \coordinate (JTL) at ({\rleg*cos(135)-1.5*\ext}, {\rleg*sin(135)+\ext*0.707});
    \coordinate (JBR) at ({\rleg*cos(-45)+1.5*\ext}, {\rleg*sin(-45)-\ext*0.707});
    \coordinate (JBL) at ({\rleg*cos(225)-1.5*\ext}, {\rleg*sin(225)-\ext*0.707});

    \draw[\mthick, Virtual] (TR) to[out=45,  in=180] (JTR);
    \draw[\mthick, Virtual] (JTR) -- ++({\ext},0);                          
    \draw[\mthick, dashed]          (JTR) to[out=45,  in=270] ++({ \ext*0.707}, {4.*\ext}); 
    \draw[\mthick]  (JTR) to[out=135, in=270] ++({-\ext*0.707}, {4.*\ext}) ; 
    \filldraw (JTR) circle[radius=\rdot];
    \draw[\mthick, Virtual] ({\rleg*cos(135)-1.5*\ext},{\rleg*sin(45)+3*\ext}) -- ({\rleg*cos(45)+3.0*\ext},{\rleg*sin(45)+3*\ext});
     \op{{\rleg*cos(45)+1.5*\ext-0.707*\ext},  {\rleg*sin(45)+3*\ext}}{0.15}{$A$};

    \draw[\mthick, Virtual] (TL) to[out=135, in=0]   (JTL) ;

    \draw[\mthick, Virtual] (BR) to[out=-45, in=180] (JBR);
    \draw[\mthick, Virtual] (JBR) -- ++({\ext},0);                           
    \draw[\mthick, dashed]          (JBR) to[out=-45, in=90] ++({ \ext*0.707},{-4.*\ext}); 
    \draw[\mthick]  (JBR) to[out=-135, in=90] ++({-\ext*0.707},{-4.*\ext}); 
    \filldraw (JBR) circle[radius=\rdot];

    \filldraw (JTR) circle[radius=\rdot];
    \draw[\mthick, Virtual] ({\rleg*cos(135)-1.5*\ext},{-\rleg*sin(45)-3*\ext}) -- ({\rleg*cos(45)+3.0*\ext},{-\rleg*sin(45)-3*\ext});
     \op{{\rleg*cos(45)+1.5*\ext-0.707*\ext},  {-\rleg*sin(45)-3*\ext}}{0.15}{$A$};

    \draw[\mthick, Virtual] (BL) to[out=225, in=0]   (JBL) ;
\end{tikzpicture}
    \end{array}
\end{equation}
with MPO tensor $A$ generating $CZX^{(N)}$. One can show that $E_2(\lambda)$ is continuous at the $\lambda=0$ point. Define the matrix product CP map $T(\lambda):=O^{(N)}(E(\lambda))$ as the horizontal contraction of $N$ copies of $E(\lambda)$ with periodic boundary condition (as~\cref{eqn:graph-O}). Since $E$ has the direct sum structure, $T(\lambda)$ is the summation
\begin{equation}
    T(\lambda) = T_1(\lambda)+T_2(\lambda)
\end{equation}
with $T_a(\lambda):=O^{(N)}(E_a(\lambda))$ for $a=1,2$. By construction,
\begin{equation}
    T_2(\lambda)=(X^{\otimes N}\otimes  \bo^{\otimes N})T_1(\lambda)(X^{\otimes N}\otimes \bo^{\otimes N})
\end{equation}
for $0\leq\lambda\leq 1$, and
\begin{equation}
\label{eqn:T2-T1-negative}
    T_2(\lambda)=(CZX^{(N)}\otimes\bo^{\otimes N})T_1(\lambda)(CZX^{(N)}\otimes\bo^{\otimes N}).
\end{equation}
for $-1\leq\lambda\leq 0$. 
In~\cref{app:explicit}, we show that $E(\lambda)$ can be constructed as the transfer matrix from a PEPS tensor $A(\lambda)$~\cite{xu2018tensor}, from which the continuity of $E_2$ at $\lambda=0$ manifests explicitly. Another way to show the continuity is by noting the symmetry at $\lambda=0$
\begin{equation}
   T_1(0)= (CZ^{(N)}\otimes  \bo^{\otimes N})T_1(0)(CZ^{(N)}\otimes \bo^{\otimes N})
\end{equation}
where $CZ^{(N)}:=\prod_{i=1}^N (CZ)_{i,i+1}$.

With the explicit form of the CP map $T(\lambda)$, we are ready to prove some properties of its fixed point $\rho(\lambda)$. First, define $\Lambda(\lambda)$ as the fixed point of $T_1(\lambda)$~\cite{wolf2012quantum}, namely, the eigenstate with eigenvalue 1 (leading eigenstate),
\begin{equation}
\begin{aligned}
    T_1(\lambda)[\Lambda(\lambda)]&=\begin{array}{c}
        \begin{tikzpicture}[scale=1., baseline={([yshift=-0.65ex] current bounding box.center)}]
         \pgfmathsetmacro{\rleg}{0.35}
    \pgfmathsetmacro{\ext}{0.2}
    \pgfmathsetmacro{\x}{2*\rleg*cos(45)+3.0*\ext}
        \opE{(0,0)};
        \opE{(\x,0)};
        \opE{(2*\x,0)};
        \opE{(3*\x,0)};
        \opsc{(1.5*\x+0.5,-1)}{2*\x}{0.2}{$\Lambda(\lambda)$}{white}
        \end{tikzpicture}
    \end{array}\\
    &=\vspace{0.5cm}\begin{array}{c}
        \begin{tikzpicture}[scale=0.9,baseline={([yshift=-0.65ex] current bounding box.center)}]
      \pgfmathsetmacro{\rleg}{0.35}
    \pgfmathsetmacro{\ext}{0.2}
    \pgfmathsetmacro{\x}{2*\rleg*cos(45)+3.0*\ext}  
        \def\off{0.4};
        \def\brastyle{dashed};
        \foreach \i in {0,1,2,3} {
          \draw[](\i*\x,-1) -- (\i*\x,-0.4);
          \draw[\brastyle](\i*\x+\off,-1) -- (\i*\x+\off,-0.4);
        }
        \opsc{(1.5*\x+0.5*\off,-1)}{2*\x}{0.2}{$\Lambda(\lambda)$}{white}
         \end{tikzpicture}.
    \end{array}
\end{aligned}
\end{equation}
The solid lines of $\Lambda$ denote the ket indices and the dashed lines denote the bra indices. 
Numerically, we show that $\Lambda(\lambda)$ is the unique fixed point and that it has finite Markov length. Due to the $\delta$ function in $E_1$, any eigenstate $w_i$ of $T_1$ with non-zero eigenvalue must be diagonal; and in particular, $\Lambda(\lambda)$ must be a diagonal density matrix. Next, at the renormalization fixed point $\lambda=\pm 1$, all components of $O_D$ are 1, from which one can show $T_1$ has rank one and $\Lambda(1)=\frac{1}{2^N}\bo^{\otimes N}$ is the only eigenstate with non-zero eigenvalue. Furthermore, $O_D(\lambda)$ has the property 
\[
\sum_{i_1 i_2 i_3 i_4 }(O_D)_{i_1 i_2 i_3 i_4}X_{i_1 j_1} X_{i_2 j_2} X_{i_3 j_3} X_{i_4 j_4} =(O_D)_{j_1 j_2 j_3 j_4},
\]
from which one can prove $[X^{\otimes N}\otimes X^{\otimes N}, T_1(\lambda)]=0$, where the first $X^{\otimes N}$ acts on the real black lines (ket indices) and the second on the dashed black lines (bra indices). Since $\Lambda(\lambda)$ is the unique leading eigenstate, it has the same symmetry
\begin{equation}
    [X^{\otimes N},\Lambda(\lambda)]=0. 
\end{equation}

We now turn to the properties of the fixed point of $T(\lambda)$. First consider the region $-1\leq \lambda\leq 0$. 
By construction, $T_1$ is Hermitian and thus allows eigendecomposition 
\[
T_1=\sum_{i:e_i\neq 0} e_i|w_i\rangle\langle w_i|,
\]
where $|w_i\rangle$ is $w_i$ reshaped into a vector. 
Therefore, using~\cref{eqn:T2-T1-negative}, 
\[
\begin{aligned}
    &T_2=\\
    &\sum_{i:e_i\neq 0} e_i (CZX^{(N)}\otimes\bo^{\otimes N})|w_i\rangle\langle w_i|(CZX^{(N)}\otimes\bo^{\otimes N}).
\end{aligned}
\]
Since $w_i$ is diagonal, 
\[
\langle w_i|(CZX^{(N)}\otimes\bo^{\otimes N})|w_j\rangle=0\quad \forall i,j:e_i\neq 0,e_j\neq 0.
\]
One concludes that the eigenstates of $T(\lambda)$ with non-zero eigenvalues are $\{|w_i\rangle,(CZX^{(N)}\otimes\bo^{\otimes N})|w_i\rangle\}$. Indeed, there are no more non-zero eigenvalues by counting the rank of $T$. In particular, the leading eigenvalue is two-fold degenerate, and the corresponding eigenstates are 
\begin{equation}
    \rho_\pm(\lambda) = (\bo^{\otimes N}\pm CZX^{(N)})\Lambda(\lambda). 
\end{equation}
We identify $\rho(\lambda)$ to be $\rho_+(\lambda)$ as it has the MPO symmetry $CZX^{(N)}\rho(\lambda)=\rho(\lambda)$. Similarly, in the region $0\leq \lambda\leq 1$, we have $\rho(\lambda)=(\bo^{\otimes N}+X^{\otimes N})\Lambda(\lambda)$.



To summarize, under this construction, $\rho(\lambda)$ takes the form of
\begin{equation}
\label{eqn:rho-form}
    \rho(\lambda)=\begin{cases}
        (\bo^{\otimes N}+X^{\otimes N}) \Lambda(\lambda), & \lambda\geq 0\\
        (\bo^{\otimes N}+CZX^{(N)}) \Lambda(\lambda), & \lambda\leq 0.
    \end{cases}
\end{equation}
where $\Lambda(\lambda)$ is a density matrix satisfying
\begin{equation}
\label{eqn:rho-property}
    \begin{aligned}
    &\Lambda(\lambda)\text{ is diagonal}\\
       &\Lambda(1)=\frac{1}{2^N}\bo^{\otimes N}\\
        &\Lambda(-\lambda)=\Lambda(\lambda)\\
        &[X^{\otimes N},\Lambda(\lambda) ]=0.
    \end{aligned}
\end{equation}
Since $\Lambda(\lambda)$ is diagonal, $[X^{\otimes N},\Lambda(\lambda) ]=0$ also leads to $[CZX^{(N)},\Lambda(\lambda) ]=0$.

\section{Phase equivalence}
Given the model $\rho(\lambda)$, in this section we prove the phase equivalence of the region $-1\leq \lambda<0$ and in the region $0<\lambda\leq 1$, given the promise that $\Lambda(\lambda)$ has a decaying CMI with a finite Markov length (see numerical input in~\cref{app:decay-CMI}).

\begin{prop}
Let $\rho(\lambda)$, for $-1 \leq \lambda \leq 1$, be a path of mixed states satisfying the properties in~\cref{eqn:rho-form,eqn:rho-property}, with the promise that $\Lambda(\lambda)$ has a finite Markov length for all $0 < \lambda \leq 1$. Then $\rho(\lambda)$ lies in the same phase (the nontrivial phase) as $\rho_\ds$ for $-1 \leq \lambda < 0$, and in the same phase (the trivial phase) as $\rho_\tc$ for $0 < \lambda \leq 1$, according to~\cref{def:phase-equiv}.
\end{prop}

We will prove the proposition in four steps. We will construct finite-depth strictly-local channel circuits $\mathcal{E},\tE$ that preserve the corresponding MPO symmetries, and then construct the low-depth quasi-local recovery channels $\mathcal{R}$ and $\tR$. 

\subsection{Step 1: $\rho(\lambda)$ to $\rho_\tc$ }
    The goal is to construct the channel circuit $\mathcal{E}$ that brings $\rho(\lambda)$ to $\rho_\tc$ for $0<\lambda\leq 1$. To start, note that the parent Lindbladian construction~\cite{liu2026parent} produces a two-site local channel $\mathcal{E}_i$ acting on site $i$ and $i+1$,
    \begin{equation}
    \label{eqn:local-E-trivial}
    \begin{aligned}
         \mathcal{E}_i(\tau)&=
         \frac{1}{4}[\tr_{i,i+1}(\tau )\otimes\bo_i\otimes \bo_{i+1} \\
         &\quad + \tr_{i,i+1}(\tau X_i\otimes X_{i+1})\otimes X_i\otimes X_{i+1}].
    \end{aligned}
\end{equation}
By construction, the local channel $\mathcal{E}_i$ has the following symmetry
\begin{equation}
\label{eqn:sym-E-X}
\begin{aligned}
    X^{\otimes N}\mathcal{E}_i(\tau) &=  \mathcal{E}_i(X^{\otimes N} \tau)\\
    \mathcal{E}_i(\tau)X^{\otimes N} &=  \mathcal{E}_i (\tau X^{\otimes N}).
\end{aligned}
\end{equation}
Now construct the channel circuit as $\mathcal{E}=\prod_{i=1}^N\mathcal{E}_i$. Since $[\mathcal{E}_i,\mathcal{E}_j]=0$ for any $i,j$, one can rewrite $\mathcal{E}=\mathcal{E}_{\mathrm{even}}\circ\mathcal{E}_{\mathrm{odd}}$ where $\mathcal{E}_{\mathrm{even}}=\prod_{i=1}^{N/2} \mathcal{E}_{2i}$ and $\mathcal{E}_{\mathrm{odd}}=\prod_{i=1}^{N/2} \mathcal{E}_{2i-1}$. Using the symmetry of local channel $\mathcal{E}_i$, one can show the channel circuit $\mathcal{E}$ has the same symmetry 
\begin{equation}
\begin{aligned}
    X^{\otimes N}\mathcal{E}(\tau) &=  \mathcal{E}(X^{\otimes N} \tau)\\
    \mathcal{E}(\tau)X^{\otimes N} &=  \mathcal{E} (\tau X^{\otimes N}).
\end{aligned}
\end{equation}

Since $\Lambda(\lambda)$ is diagonal, one can prove $\tr_{i,i+1}(\Lambda(\lambda) X_i\otimes X_{i+1})=0$, and therefore 
\begin{equation}
    \mathcal{E}_{\mathrm{odd}}(\Lambda)=\frac{1}{4^{N/2}}\bo^{\otimes N}, 
\end{equation}
and $\mathcal{E}(\Lambda)=\frac{1}{2^N}\bo^{\otimes N}$. Using the symmetry of the channel circuit,
\begin{equation}
\begin{aligned}
    \mathcal{E}(\rho(\lambda))&=\mathcal{E}((\bo^{\otimes N}+X^{\otimes N})\Lambda(\lambda))\\
    &=(\bo^{\otimes N}+X^{\otimes N}) \mathcal{E}(\Lambda(\lambda))\\
    &=\frac{1}{2^N}(\bo^{\otimes N}+X^{\otimes N})=\rho_\tc. 
\end{aligned}
\end{equation}
$\mathcal{E}$ is a depth-2 local quantum circuit. This finishes step 1.

Before moving on, we consider the action of some local channels $\{\mathcal{E}_i\}$ on $\Lambda(\lambda)$, and denote 
\begin{equation}
    \Lambda^{[S]}(\lambda):=\left(\prod_{i\in S} \mathcal{E}_i\right) \Lambda(\lambda)
\end{equation}
where $S\subseteq \{1,2,\cdots N\}$. The symmetries~\cref{eqn:sym-E-X} and $[\Lambda(\lambda),X^{\otimes N}]=0$ lead to
\begin{equation}
    [\Lambda^{[S]}(\lambda),X^{\otimes N}]=0,
\end{equation}
that $\Lambda^{[S]}(\lambda)$ also has the symmetry $X^{\otimes N}$. Together with the fact that $\Lambda(\lambda)$ is diagonal and $\Lambda^{[S]}(\lambda)$ is also diagonal, its reduced density matrix on region $A$, $\Lambda_A^{[S]}(\lambda):=\tr_{\bar{A}}\Lambda^{[S]}(\lambda) $ also has the symmetry
\begin{equation}
\label{eqn:reduced-rho-sym}
    [\Lambda_A^{[S]}(\lambda), X^{\otimes N_A}]=0, 
\end{equation}
where $N_A$ denote the size of region $A$. This symmetry condition is crucial in the next step.

\begin{figure*}[t]
  \centering
  \begin{tikzpicture}[
    gate/.style={draw, fill=blue!15, minimum width={\gs*1.4 cm}, minimum height=\gateH, inner sep=0pt, rounded corners=1pt},
    wire/.style={thick},
    >=Stealth,
    rvgate/.style={draw, fill=blue!25, minimum width={\gs*5.0 cm}, minimum height=\gateH, inner sep=0pt, rounded corners=1pt},
    wire/.style={thick},
    >=Stealth,
    gatew/.style n args={1}{draw, fill=blue!25, minimum width={#1*\gs*1.0 cm}, minimum height=\gateH, inner sep=0pt, rounded corners=1pt},
    wire/.style={thick},
    >=Stealth
  ]
  \pgfmathsetmacro{\nwires}{18}       
  \pgfmathsetmacro{\gs}{0.3}          
  \pgfmathsetmacro{\ls}{1.0}          
  \def\gateH{0.2cm}                   
  \pgfmathsetmacro{\wireExt}{0.4}     
  \pgfmathsetmacro{\nlayers}{2}       
  \pgfmathsetmacro{\sep}{0.4}         

  \pgfmathsetmacro{\maxIdx}{int(\nwires-1)}
  \pgfmathsetmacro{\circuitWidth}{(\nwires-1)*\gs}
  \pgfmathsetmacro{\offsetB}{\circuitWidth + \sep}  
  
  \pgfmathsetmacro{\labelY}{1.4*\nlayers*\ls + \wireExt + 0.5}
    \pgfmathsetmacro{\midX}{0.5*(\nwires-1)*\gs}

  \begin{scope}[shift={(0,0.5*\ls)},scale=0.9]
    \foreach \i in {0,...,\maxIdx} {
      \pgfmathsetmacro{\xpos}{\i*\gs}
      \draw[wire] (\xpos, \ls-\wireExt) -- (\xpos, {\nlayers*\ls + \wireExt});
    }
    \pgfmathsetmacro{\yA}{\ls}
    \pgfmathsetmacro{\maxPairA}{int(\nwires/2 - 1)}
    \foreach \k in {0,...,\maxPairA} {
      \pgfmathsetmacro{\xmid}{(2*\k+0.5)*\gs}
      \node[gate] at (\xmid, \yA) {};
    }
    \pgfmathsetmacro{\yB}{2*\ls}
    \pgfmathsetmacro{\maxPairB}{int((\nwires-1)/2)}
    \foreach \k in {-1,...,\maxPairB} {
      \pgfmathsetmacro{\xmid}{(2*\k+1.5)*\gs}
      \node[gate] at (\xmid, \yB) {};
    }
    \pgfmathsetmacro{\coverLeft}{-\gs*2.0}
    \pgfmathsetmacro{\coverBot}{\yB-0.3}
    \pgfmathsetmacro{\coverTop}{\yB+0.3}
    \pgfmathsetmacro{\coverright}{-\gs*0.5}
    \fill[white] (\coverLeft, \coverBot) rectangle (\coverright, \coverTop);
    \pgfmathsetmacro{\coverRight}{(\nwires-1)*\gs + \gs*2.0}
    \pgfmathsetmacro{\rightEdge}{(\nwires-0.5)*\gs}
    \fill[white] (\rightEdge, \coverBot) rectangle (\coverRight, \coverTop);
  \end{scope}

  \begin{scope}[shift={(\offsetB,0)},scale=0.9]
    \foreach \i in {0,...,\maxIdx} {
      \pgfmathsetmacro{\xpos}{\i*\gs}
      \draw[wire] (\xpos, \ls-\wireExt) -- (\xpos, {1.5*\nlayers*\ls + \wireExt});
    }
    \pgfmathsetmacro{\yA}{\ls}
    \pgfmathsetmacro{\maxPairA}{int(\nwires/2 - 1)}
    \foreach \k in {0,...,\maxPairA} {
      \pgfmathsetmacro{\xmid}{(2*\k+0.5)*\gs}
      \pgfmathsetmacro{\rem}{int(mod(\k,3))}
      \pgfmathsetmacro{\ypos}{\yA*(1 + \rem/3)}
      \node[gate] at (\xmid, \ypos) {};
    }
    \pgfmathsetmacro{\yB}{2*\ls}
    \pgfmathsetmacro{\maxPairB}{int((\nwires-1)/2 )}
    \foreach \k in {-1,...,\maxPairB} {
      \pgfmathsetmacro{\xmid}{(2*\k+1.5)*\gs}
      \pgfmathsetmacro{\rem}{int(mod(\k+3,3))}
      \pgfmathsetmacro{\ypos}{1.2*\yB+\ls*\rem/3}
      \node[gate] at (\xmid, \ypos) {};
    }
    \pgfmathsetmacro{\coverLeft}{-\gs*2.0}
    \pgfmathsetmacro{\coverBot}{1.2*\yB-0.3+\ls*2/3}
    \pgfmathsetmacro{\coverTop}{1.2*\yB+0.3+\ls*2/3}
    \pgfmathsetmacro{\coverright}{-\gs*0.5}
    \fill[white] (\coverLeft, \coverBot) rectangle (\coverright, \coverTop);
    \pgfmathsetmacro{\coverRight}{(\nwires-1)*\gs + \gs*2.0}
    \pgfmathsetmacro{\rightEdge}{(\nwires-0.5)*\gs}
    \fill[white] (\rightEdge, \coverBot) rectangle (\coverRight, \coverTop);

    \pgfmathsetmacro{\annoGateX}{(3.5)*\gs}
    \pgfmathsetmacro{\annoGateBot}{\yA - 0.5}
    \pgfmathsetmacro{\gateHalfW}{\gs*5.0*0.5}  
    \draw[<->, thick] ({\annoGateX - \gateHalfW}, \annoGateBot) -- ({\annoGateX + \gateHalfW}, \annoGateBot)
      node[midway, below, font=\small] {$2r$};
  \end{scope}

  \begin{scope}[shift={(2*\offsetB,0)},scale=0.9]
    \foreach \i in {0,...,\maxIdx} {
      \pgfmathsetmacro{\xpos}{\i*\gs}
      \draw[wire] (\xpos, \ls-\wireExt) -- (\xpos, {1.5*\nlayers*\ls + \wireExt});
    }
    \pgfmathsetmacro{\yA}{\ls}
    \pgfmathsetmacro{\maxPairA}{int(\nwires/2 - 1)}
    \foreach \k in {0,...,\maxPairA} {
      \pgfmathsetmacro{\xmid}{(2*\k+0.5)*\gs}
      \pgfmathsetmacro{\rem}{int(mod(\k,3))}
      \pgfmathsetmacro{\ypos}{\yA*(1 + \rem/3)}
      \node[rvgate] at (\xmid, \ypos) {};
    }
    \node[gatew={3.0}] at (17.5*\gs, \yA) {};
    \node[gatew={3.0}] at (-0.5*\gs, \yA*5/3) {};
    \pgfmathsetmacro{\yB}{2.4*\ls}
    \pgfmathsetmacro{\maxPairB}{int((\nwires-1)/2 - 1)}
    \foreach \k in {0,...,\maxPairB} {
      \pgfmathsetmacro{\xmid}{(2*\k+1.5)*\gs}
      \pgfmathsetmacro{\rem}{int(mod(\k,3))}
      \pgfmathsetmacro{\ypos}{\yB+\ls*\rem/3}
      \node[rvgate] at (\xmid, \ypos) {};
    }
    \node[gatew={1.5}] at (17.5*\gs, \yB) {};
    \node[gatew={3.0}] at (0.5*\gs, \yB+\ls*2/3) {};
    \node[gatew={3.0}] at (16.5*\gs, \yB+\ls*2/3) {};
    \pgfmathsetmacro{\coverLeft}{-\gs*3.0}
     \pgfmathsetmacro{\coverright}{-\gs*0.5}
    \pgfmathsetmacro{\coverBot}{\yA-0.3}
    \pgfmathsetmacro{\coverTop}{\yB+0.3+\ls*2/3}
    \fill[white] (\coverLeft, \coverBot) rectangle (\coverright, \coverTop);
    \pgfmathsetmacro{\coverRight}{(\nwires-1)*\gs + \gs*3.0}
    \pgfmathsetmacro{\rightEdge}{(\nwires-0.5)*\gs}
    \fill[white] (\rightEdge, \coverBot) rectangle (\coverRight, \coverTop);
    \pgfmathsetmacro{\annoGateX}{(6.5)*\gs}
    \pgfmathsetmacro{\annoGateBot}{\yA - 0.5}
    \pgfmathsetmacro{\gateHalfW}{\gs*5.0*0.5}  
    \draw[<->, thick] ({\annoGateX - \gateHalfW}, \annoGateBot) -- ({\annoGateX + \gateHalfW}, \annoGateBot)
      node[midway, below, font=\small] {$2r$};

    \pgfmathsetmacro{\annoX}{(\nwires-0.5)*\gs + 0.4}
    \pgfmathsetmacro{\depthBot}{\yA}
    \pgfmathsetmacro{\depthTop}{\yB + \ls*2/3}
    \draw[<->, thick] (\annoX, \depthBot) -- (\annoX, \depthTop)
      node[midway, right, font=\small] {$2r$};
  \end{scope}
    \node[font=\large] at (\midX, \labelY) {$\mathcal{E}$};
    \node[] at (0, \labelY) {(a)};

    \node[font=\large] at ({\offsetB + \midX}, \labelY) {$\mathcal{E}$ (reorganized)};
    \node[] at (\offsetB, \labelY) {(b)};

    \node[font=\large] at ({2*\offsetB + \midX}, \labelY) {$\mathcal{R}$};
    \node[] at (2*\offsetB, \labelY) {(c)};

  \end{tikzpicture}
  \caption{Configurations of channel circuit (reproduce from~\cite{sang2025stability}). (a) The original local finite-depth channel circuit $\mathcal{E}$. (b) The reorganized circuit is such that two channels in each layer are separated by $2r$. (c) The recovery channel circuit $\mathcal{R}$, which is quasi-local and low-depth. }
  \label{fig:circuit}
\end{figure*}
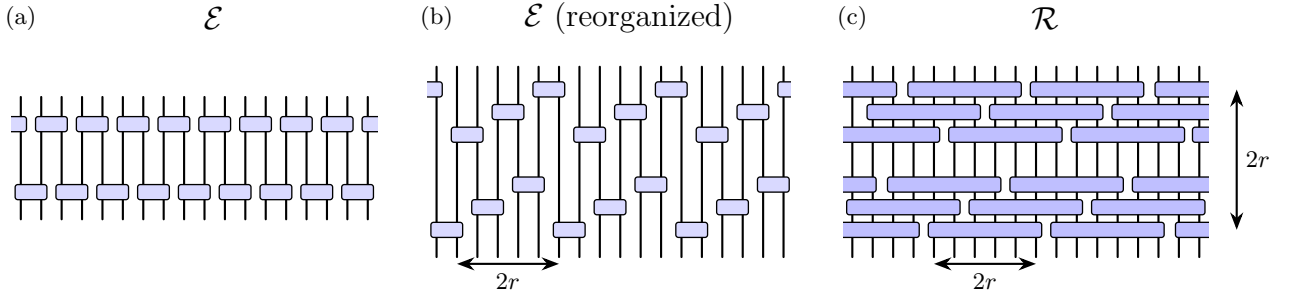

\subsection{Step 2: $\rho_\tc$ to $\rho(\lambda)$} 
The goal of this step is to construct the channel circuit $\mathcal{R}$ that brings $\rho_\tc$ to $\rho(\lambda)$ for $0<\lambda\leq 1$. We first show the channel circuit that brings $\Lambda(1)=\bo^{\otimes N}/2^N$ to $\Lambda(\lambda)$, then prove that the same channel circuit is able to bring $\rho_\tc$ to $\rho(\lambda)$.

Under the promise that $\Lambda(\lambda)$ has a finite Markov length, the channel circuit $\mathcal{R}$ that brings $\Lambda(1)=\bo^{\otimes N}/2^N$ to $\Lambda(\lambda)$ exists by~\cref{prop:reverse},  with error controlled by $\epsilon$. To construct $\mathcal{R}$ explicitly, we follow the procedure of~\cite{sang2025stability} to first reorganize the channel circuit $\mathcal{E}$ such that two channels within the same layer are separated from each other by a distance $2r$, and the number of layers is increased from 2 to $2r$ (see Fig.~\ref{fig:circuit}). Label the channels in the reorganized circuit as $\mathcal{E}_{l,x}$ where $l=1,\cdots,2r$ and $x=1,\cdots,N/2r$. One can then construct a reverse channel circuit $\mathcal{R}=\prod_{l,x}\mathcal{R}_{l,x}$ with $2r$ layers and $\mathrm{supp}(\mathcal{R}_{l,x})=2r$.

Specifically, the explicit form of $\mathcal{R}_{l,x}$ is given by the Petz recovery map. Consider the action of $\mathcal{E}_{l,x}$ supported on site $i,i+1$ for some $i$. The state it acts on is $\Lambda^{[S_{l,x}]}(\lambda)$, where the set $S_{l,x}$ is
\[
\begin{aligned}
    &S_{l,x}=\\
    &\{(l',x')|1\leq l'<l, 1\leq x'\leq \frac{N}{2r}\}\cup\{(l,x')|1\leq x'<x\},
\end{aligned}
\]
namely, $\Lambda^{[S_{l,x}]}(\lambda)$ is obtained by acting channels to the left and below $\mathcal{E}_{l,x}$ to $\Lambda(\lambda)$. Then, $\mathcal{E}_{l,x}(\Lambda^{[S_{l,x}]}(\lambda))$ can be approximated recovered by $\mathcal{R}_{l,x}$. Denote region $A=\{i,i+1\}$, and region $B$ as the region surrounding $A$ of width $r$, $B=\{i-r,i-1\}\cup\{i+2,i+r+1\}$, and $C=\mathcal{L}\setminus (A\cup B)$. Then $\mathcal{R}_{l,x}$ is supported only on $A\cup B$ and given explicitly by the twirled Petz map~\cite{petz1986sufficient,junge2018universal,sang2025stability}:
\begin{widetext}
\begin{equation}
\label{eqn:local-reverse}
    \mathcal{R}_{l,x}(\cdot)=\int_{-\infty}^{\infty} f(t) (\Lambda^{[S_{l,x}]}_{AB})^{\frac{1-it}{2}} \mathcal{E}_{l,x}^{\dg}\left(\mathcal{E}_{l,x}(\Lambda^{[S_{l,x}]}_{AB})^{\frac{-1+it}{2}}(\cdot) \mathcal{E}_{l,x}(\Lambda^{[S_{l,x}]}_{AB})^{\frac{-1-it}{2}}\right) (\Lambda^{[S_{l,x}]}_{AB})^{\frac{1+it}{2}} dt
\end{equation}
\end{widetext}
where $f(t):=\frac{\pi}{2(\cosh(\pi t)+1)}$, and $\Lambda_{AB}^{[S_{l,x}]}(\lambda)=\tr_{AB}(\Lambda^{[S_{l,x}]}(\lambda))$. We omit $\lambda$ in the above equation for the ease of presentation. 

We now prove that the same channel circuit $\mathcal{R}$ can bring $\rho_\tc$ to $\rho(\lambda)$ with error controlled by $\epsilon$. Using~\cref{eqn:sym-E-X,eqn:reduced-rho-sym} and the fact that $\mathcal{E}_{l,x}$ being self-dual leads to
\begin{equation}
\begin{aligned}
    X^{\otimes N}\mathcal{R}_{l,x}(\tau) &=  \mathcal{R}_{l,x}(X^{\otimes N} \tau)\\
    \mathcal{R}_{l,x}(\tau)X^{\otimes N} &=  \mathcal{R}_{l,x} (\tau X^{\otimes N}).
\end{aligned}
\end{equation}
Therefore, $\mathcal{R}=\prod_{l,x}\mathcal{R}_{l,x}$ has the same symmetry property, and
\begin{equation}
    \mathcal{R}((\bo^{\otimes N}+X^{\otimes N})\Lambda(\lambda))=(\bo^{\otimes N}+X^{\otimes N}) \mathcal{R}(\Lambda(\lambda)).
\end{equation}
The error of the recovery is then bounded by
\begin{equation}
\begin{aligned}
    &\|\mathcal{R}((\bo^{\otimes N}+X^{\otimes N})\Lambda)-\frac{1}{2^N}(\bo^{\otimes N}+X^{\otimes N})\|_1 \\
    &= \|\mathcal{R}(\Lambda)-\frac{1}{2^N}\bo^{\otimes N}\|_1+\|X^{\otimes N}\mathcal{R}(\Lambda)-\frac{1}{2^N}X^{\otimes N}\|_1\\
    &\leq \|\mathcal{R}(\Lambda)-\frac{1}{2^N}\bo^{\otimes N}\|_1+ \|X^{\otimes N}\|_\infty \|\mathcal{R}(\Lambda)-\frac{1}{2^N}\bo^{\otimes N}\|_1\\
    &\leq 2 \|\mathcal{R}(\Lambda)-\frac{1}{2^N}\bo^{\otimes N}\|_1 \leq 2\epsilon,
\end{aligned}
\end{equation}
where we use triangle inequality for Schatten norm, $\|AB\|_1\leq \|A\|_1 \|B\|_\infty$, and $\|X^{\otimes N}\|_\infty=1$. By choosing $r=\xi\log(\mathrm{poly}N)$, the channel $\mathcal{R}$ is low-depth quasi-local, and the error $\epsilon$ vanishes in the thermodynamic limit $N\ra\infty$. This finishes step 2. 

\subsection{Step 3: $\rho(\lambda)$ to $\rho_\ds$} 
The goal is to construct the channel circuit $\tE$ that brings $\rho(\lambda)$ to $\rho_\ds$ for $-1\leq \lambda <0$. A valid construction comes from the parent Lindbladian construction, which produces a four-site channel $\tE_i$ acting on site $i,i+1,i+2,i+3$ (see~\cite{liu2026parent} Appendix F), 
\begin{equation}
\label{eqn:CZX-channel-1}
    \tE_i(\tau)=\frac{1}{16}\sum_{m=1}^8 \tr[B_m^\dg \tau] B_m
\end{equation}
with 
\begin{equation}
\label{eqn:CZX-channel-2}
    \begin{aligned}
        B_1 & = \bo_2\otimes \bo_2\otimes \bo_2\otimes \bo_2, B_2=Z\otimes \bo_2\otimes \bo_2\otimes \bo_2\\
        B_3&=\bo_2\otimes \bo_2\otimes \bo_2\otimes Z, B_4=Z\otimes \bo_2\otimes \bo_2\otimes Z\\
        B_5&= CZX^{(4)}(\bo_2\otimes \bo_2\otimes \bo_2\otimes\bo_2) \\
        B_6&=CZX^{(4)} (Z\otimes \bo_2\otimes \bo_2\otimes \bo_2)\\
        B_7&= CZX^{(4)}(\bo_2\otimes \bo_2\otimes \bo_2\otimes Z)\\
        B_8&=CZX^{(4)} (Z\otimes \bo_2\otimes \bo_2\otimes Z)\\
    \end{aligned}.
\end{equation}
Note that $\tE_i$ is supported on 4 sites instead of 2 sites, because $\rho_\ds$ is at the renormalization fixed point after blocking once. 
Denote $\tE=\prod_{i=1}^{N/2} \tE_{2i-1}$, and since the local channels commute $[\tE_i,\tE_{i+2}]=0$~\cite{liu2026parent}, $\tE$ is again a depth-2 strictly-local channel circuit. To show that $\tE$ brings $\rho(\lambda)$ to $\rho_\ds$, we first note that $\tE(\Lambda(\lambda))=\frac{1}{2^N}\bo^{\otimes N}$ since $\Lambda(\lambda)$ is diagonal. Next, we show that $\tE_i$ has the global $CZX^{(N)}$ symmetry. This can be proven using the MPO representation $A$ of $CZX^{(N)}$, 
\begin{equation}
    A^{01}=\begin{pmatrix}
        1 & 1 \\
        0 & 0
    \end{pmatrix}\quad A^{10}=\begin{pmatrix}
        0 & 0 \\
        1 & -1
    \end{pmatrix}.
\end{equation}
Denote $A^{(4)}$ as the tensor obtained by horizontally concatenating four copies of $A$, graphically,
\begin{equation}
\label{eqn:A4}
   A^{(4)}_{\alpha\beta}:=
    \begin{array}{c}
\begin{tikzpicture}[scale=1]
    \pgfmathsetmacro{\x}{0.494}       
    \pgfmathsetmacro{\dx}{0.988}      
    \pgfmathsetmacro{\vl}{0.247}      
    \pgfmathsetmacro{\yo}{0.564}      
    \pgfmathsetmacro{\ybot}{0.318}    
    \pgfmathsetmacro{\ytop}{1.129}    
    \pgfmathsetmacro{\yleg}{0.811}    

    \draw[Virtual] (\x+\vl,        \yo) -- (\x+\dx-\vl,      \yo); 
    \draw[Virtual] (\x+\dx+\vl,    \yo) -- (\x+2*\dx-\vl,    \yo); 
    \draw[Virtual] (\x+2*\dx+\vl,  \yo) -- (\x+3*\dx-\vl,    \yo); 

    \draw[Virtual] (\x-\vl-0.3*\dx,        \yo) -- (\x,               \yo); 
    \draw[Virtual] (\x+3*\dx,      \yo) -- (\x+3*\dx+\vl+0.3*\dx,     \yo); 

    \node[anchor=east]  at (\x-\vl-0.3*\dx,       \yo) {$\alpha$};           
    \node[anchor=west]  at (\x+3*\dx+\vl+0.3*\dx, \yo) {$\beta$};            

    \draw[-mid] (\x,        \yleg) -- (\x,        \ytop);
    \draw[-mid] (\x+\dx,    \yleg) -- (\x+\dx,    \ytop);
    \draw[-mid] (\x+2*\dx,  \yleg) -- (\x+2*\dx,  \ytop);
    \draw[-mid] (\x+3*\dx,  \yleg) -- (\x+3*\dx,  \ytop);

    \draw[bevel, -mid] (\x,        0) -- (\x,        \ybot);
    \draw[bevel, -mid] (\x+\dx,    0) -- (\x+\dx,    \ybot);
    \draw[bevel, -mid] (\x+2*\dx,  0) -- (\x+2*\dx,  \ybot);
    \draw[bevel, -mid] (\x+3*\dx,  0) -- (\x+3*\dx,  \ybot);

    \Osymb{(\x,        \yo)};
    \Osymb{(\x+\dx,    \yo)};
    \Osymb{(\x+2*\dx,  \yo)};
    \Osymb{(\x+3*\dx,  \yo)};
\end{tikzpicture}
        \end{array}
        \,.
\end{equation}
One can verify that for all four choices of $\alpha,\beta=0,1$ and arbitrary $\tau$, $\tE_i(A_{\alpha\beta}^{(4)} \tau)=A_{\alpha\beta}^{(4)} \tE_i(\tau)$ and $\tE_i(\tau A_{\alpha\beta}^{(4)})=\tE_i(\tau) A_{\alpha\beta}^{(4)}$, 
and subsequently,
\begin{equation}
\label{eqn:step-3-commute-local}
    \begin{aligned}
        &\tE_i(CZX^{(N)}\tau)=CZX^{(N)}\tE_i(\tau)\\
        &\tE_i(\tau CZX^{(N)})=\tE_i(\tau) CZX^{(N)}.
    \end{aligned}
\end{equation}
We will present a generic proof in~\cref{sec:general}.  
The full channel circuit therefore has the same symmetry property, 
\begin{equation}
    \begin{aligned}
        &\tE(CZX^{(N)}\tau)=CZX^{(N)}\tE(\tau)\\
        &\tE(\tau CZX^{(N)})=\tE(\tau) CZX^{(N)}.
    \end{aligned}
\end{equation}
Combined with the fact that $\tE(\Lambda(\lambda))=\frac{1}{2^N}\bo^{\otimes N}$, we conclude
\begin{equation}
\begin{aligned}
    &\tE((\bo^{\otimes N}+CZX^{(N)})\Lambda(\lambda))\\
    =&(\bo^{\otimes N}+CZX^{(N)})\tE(\Lambda(\lambda))\\
    =&\frac{1}{2^N} (\bo^{\otimes N}+CZX^{(N)}). 
\end{aligned}
\end{equation}
This finishes step 3. 

\subsection{Step 4: $\rho_\ds$ to $\rho(\lambda)$}
The goal is to construct the channel circuit $\tR$ that brings $\rho_\ds$ to $\rho(\lambda)$ for $-1\leq \lambda <0$ with error controlled by $\epsilon$. This step is identical to step 2. Define
\begin{equation}
    \tilde{\Lambda}^{[S]}(\lambda):=\left(\prod_{i\in S} \tE_i\right) \Lambda(\lambda)
\end{equation}
where $S\subseteq \{1,3,\cdots,N-1\}$.~\cref{eqn:step-3-commute-local} and $[\Lambda(\lambda),CZX^{(N)}]=0$ leads to
\begin{equation}
    [\tilde{\Lambda}^{[S]}(\lambda),CZX^{(N)}]=0. 
\end{equation}
Since $\tilde{\Lambda}^{[S]}(\lambda)$ is also diagonal, define $CZ^{(N)}=\prod_{i=1}^N (CZ)_{i,i+1}$, $\tilde{\Lambda}^{[S]}(\lambda)$ trivially commutes with $CZ^{(N)}$ and therefore $ [\tilde{\Lambda}^{[S]}(\lambda), X^{\otimes N}]=0$. The reduced density matrix also has the same symmetry, $[\tilde{\Lambda}_A^{[S]}(\lambda)\otimes \bo^{N_A}, X^{\otimes N}]=0$, which leads to 
\begin{equation}
\label{eqn:reduced-Lambda-comm-3}
    [\tilde{\Lambda}_A^{[S]}(\lambda)\otimes \bo^{\otimes N_{\bar{A}}}, CZX^{(N)}]=0
\end{equation}
since $\tilde{\Lambda}_A^{[S]}(\lambda)$ is diagonal. 

By constructing the local channel $\tR_{l,x}$ similar to~\cref{eqn:local-reverse}, using~\cref{eqn:step-3-commute-local,eqn:reduced-Lambda-comm-3} and $\tE_{i}$ is self-adjoint leads to
\begin{equation}
\begin{aligned}
    &CZX^{(N)}\tR_{l,x}(\tau)=\tR_{l,x}( CZX^{(N)}\tau)\\
    &\tR_{l,x}(\tau)CZX^{(N)}=\tR_{l,x}( \tau CZX^{(N)}).
\end{aligned}
\end{equation}
Define $\tR=\prod_{l,x}\tR_{l,x}$, follow the same derivation as in step 2,
\begin{equation}
    \|\tR((\bo^{\otimes N}+CZX^{(N)})\Lambda)-\frac{1}{2^N}(\bo^{\otimes N}+CZX^{(N)})\|_1\leq 2\epsilon. 
\end{equation}
By choosing $r=\xi\log(\mathrm{poly}N)$, the channel $\tR$ is low-depth quasi-local, and the error $\epsilon$ vanishes in the thermodynamic limit $N\ra\infty$. This finishes step 4 and the whole proof.



\section{Generalization}
\label{sec:general}

From the proof above, we see that one important ingredient is to correctly choose the local channel $\mathcal{E}_i$ and $\tE_i$, such that the channel preserves the desired global symmetry. This makes it possible to construct the recovery channels $\mathcal{R}_i$ and $\tR_i$. 

The above model corresponds to the boundary of $\mathbb{Z}_2$ topological order (toric code and double semion). 
To generalize to boundary of $\mathbb{Z}_{\N}$ topological order, we first re-examine the local channel $\tilde{\mathcal{E}}_i$ in~\cref{eqn:CZX-channel-1,eqn:CZX-channel-2}. Using the MPO representation $A$ of $CZX^{(N)}$, one can show
\begin{equation}
    \mathrm{Span}\{A_{\alpha\beta}^{(4)}\}_{\alpha\beta}=\mathrm{Span}\{B_5,B_6,B_7,B_8\}. 
\end{equation}
By the explicit form of $B_m$, we see that $\{B_{m'}^\dg B_m\}_m$ is a permutation of $\{B_m\}_m$ for any chosen $m'$, up to a sign. Therefore, 
\begin{equation}
\begin{aligned}
    \tE_i(B_{m'} \tau)&=B_{m'} \tE_i(\tau)\\
    \tE_i( \tau B_{m'})&=\tE_i(\tau)B_{m'},\quad \forall m'=1,\cdots,8
\end{aligned}
\end{equation}
which then leads to~\cref{eqn:step-3-commute-local}. The construction of the local channel in the nontrivial region for $\mathbb{Z}_{\N}$ will generalize this observation. 

\subsection{Local channel in the nontrivial region}
In this section, we construct the local channel $\tE_i$ that generalizes~\cref{eqn:CZX-channel-1,eqn:CZX-channel-2} to $\mathbb{Z}_{\N}$. Specifically, we will construct a local channel that preserves the $\mathbb{Z}_\N$ MPO symmetry associated with a non-trivial 3-cocycle. The main result is~\cref{eqn:ZN-channel}. 

For Abelian group $\mathbb{Z}_{\N}$, we label the MPO symmetries as $\{O^{(N)}_{g}\}_{g=0,\cdots,\N-1}$ where $O_g^{(N)}$ is generated by tensor $A_g$. One particular construction is provided in~\cite{liu2025trading,GarreRubio2023classifyingphases}, with MPO tensor
\begin{equation}
\begin{aligned}
    A_g^{gh,h}&=\omega_g |h)( h|\\
  \mathrm{where}\quad  \omega_g & = \sum_{kl} \omega(g,k,l-k) |k)( l|.
\end{aligned}
\end{equation}
Here $|h)$ denotes a ket state supported on the virtual bond, and $\omega(g,h,k)\in\mathcal{H}^3(\mathbb{Z}_{\N}, U(1))$ is a 3-cocycle. For the non-trivial region, we choose
\begin{equation}
    \omega(g,h,k)=\exp\left(\frac{2\pi i g}{\N^2}([h]_\N+ [k]_\N - [h+k]_\N)\right)
\end{equation}
with $[h]_\N:=h\mod \N$. The above MPO tensor has physical dimension $d=\N$ and bond dimension $D=\N$. For the qudit system with local dimension $\N$, the generalized $X$ and $Z$ operators are defined by
\begin{equation}
\begin{aligned}
    X|j\rangle&= |[j+1]_\N\rangle \\
    Z|j\rangle&= \phi^j |j\rangle\quad \phi=e^{2\pi i/\N}. 
\end{aligned}
\end{equation}
They satisfy the commutation relation $ZX = \phi XZ$. 

Using the MPO tensors and the underlying pre-bialgebraic structure~\cite{molnar2022matrix,liu2025trading} (specifically, the coproduct definition, see a detailed discussion in~\cref{app:details}), one can show
\begin{equation}
\label{eqn:span-4-general}
\begin{aligned}
    &\mathrm{Span}\{A_{g,\alpha\beta}^{(4)}\}_{\alpha\beta}=\\
    &\quad \mathrm{Span}\{(X^g)^{\otimes 4} D_g^{(4)} (Z^{k_1}\otimes \bo\otimes \bo\otimes Z^{k_2}) \}_{k_1,k_2}
\end{aligned}
\end{equation}
where
\begin{equation}
\begin{aligned}
   & D_g^{(4)}=\frac{\omega(g,k,m-k)\omega(g,m,n-m)\omega(g,n,l-n)}{\omega(g,k,l-k)}\\
    &\quad \quad |k,m,n,l\rangle\langle k,m,n,l|. 
\end{aligned}
\end{equation}
Define the operators
\begin{equation}
    B_{g,k_1, k_2}:=(X^g)^{\otimes 4} D_g^{(4)} (Z^{k_1}\otimes \bo\otimes \bo\otimes Z^{k_2}). 
\end{equation}
In particular, $B_{0,0,0}=\bo$. 
We construct the local channel as
\begin{equation}
\label{eqn:ZN-channel}
    \tE_i(\tau)=\frac{1}{\N^4}\sum_{g,k_1,k_2=0}^{\N-1}\tr[B_{g,k_1,k_2}^\dg \tau]  B_{g,k_1,k_2}.
\end{equation}

To see that this channel preserves the MPO symmetries, using the property (see a proof in~\cref{app:details})
\begin{equation}
\label{eqn:commute-general}
    \begin{aligned}
        &[D_g^{(4)},(X^{g'})^{\otimes 4}]=0,\quad \forall g,g'
    \end{aligned}
\end{equation}
one can show
\begin{equation}
\begin{aligned}
    B_{g,k_1,k_2}^\dg B_{g',k_1',k_2'}&=B^\dg_{g-g',k_1-k_1',k_2-k_2'} \phi^{(k_1'+k_2')(g-g')}\\
    B_{g',k_1',k_2'}^\dg B_{g,k_1,k_2}&=B_{g-g',k_1-k_1',k_2-k_2'} \phi^{(k_1'+k_2')(g'-g)}.
\end{aligned}
\end{equation}
The two phases cancel and therefore
\begin{equation}
    \begin{aligned}
        &B_{g',k_1',k_2'}\tE_i(\tau)=\tE_i(B_{g',k_1',k_2'} \tau)\\
        &\tE_i(\tau)B_{g',k_1',k_2'}=\tE_i( \tau B_{g',k_1',k_2'}).
    \end{aligned}
\end{equation}
The local channel then has the symmetry property, 
\begin{equation}
    \begin{aligned}
        &\tE_i(O_g^{(N)} \tau)=O_g^{(N)}\tE_i(\tau)\\
        &\tE_i(\tau O_g^{(N)})=\tE_i(\tau) O_g^{(N)}\quad\forall g,
    \end{aligned}
\end{equation}
and the channel circuit shares the same symmetry. 

We comment that the construction relies on the pre-bialgebraic structure that goes beyond weak Hopf algebra, as analyzed in~\cite{liu2025trading}. Such a pre-bialgebraic is crucial for the local channel from the parent Lindbladian to be commuting. If we use the weak Hopf algebra instead, the local channels fail to commute, and the channel circuit $\tE$ is not low-depth. 

\subsection{Local channel in the trivial region}
In the trivial region, the local channel can be constructed simply as
\begin{equation}
    \mathcal{E}_i(\tau)=\frac{1}{\N^2}\sum_{g=0}^{\N-1}\tr(X_i^{\dg g}\otimes X^{\dg g}_{i+1} \tau ) X_i^g\otimes X^g_{i+1},
\end{equation}
which is a direct generalization of~\cref{eqn:local-E-trivial}. It preserves the $\mathbb{Z}_\N$ MPO symmetry associated with trivial 3-cocycle,
\begin{equation}
    \begin{aligned}
        &\mathcal{E}_i((X^{g'})^{\otimes N} \tau)=(X^{g'})^{\otimes N}\mathcal{E}_i(\tau)\\
        &\mathcal{E}_i(\tau (X^{g'})^{\otimes N})=\mathcal{E}_i(\tau)(X^{g'})^{\otimes N}\quad\forall g'.
    \end{aligned}
\end{equation}

\section{Phase transition point at $\lambda=0$}
Having established the phase equivalence, we now turn to the phase transition point at $\lambda_c=0$. The density matrix $\rho_c:=\rho(0)$ does not admit a matrix product density operator (MPDO) representation with finite bond dimension. Therefore, in this section we rely on numerical methods.

The property of the vectorized state $|\rho_c\rangle\!\rangle$ was studied in~\cite{xu2018tensor}, where the vectorized state $|\rho\rangle\!\rangle$ of a density matrix $\rho=\rho_{ij}|i\rangle\langle j|$ is defined as $|\rho\rangle\!\rangle:=\rho_{ij}|i\rangle\otimes |j\rangle$. By computing the entanglement entropy scaling of $|\rho_c\drangle$ and fitting with the scaling law from conformal field theory (CFT), it was found that $|\rho_c\drangle$ behaves as the ground state of compactified boson CFT with central charge $c=1$.

\begin{figure}[t]
    \centering
    \includegraphics[width=0.49\linewidth]{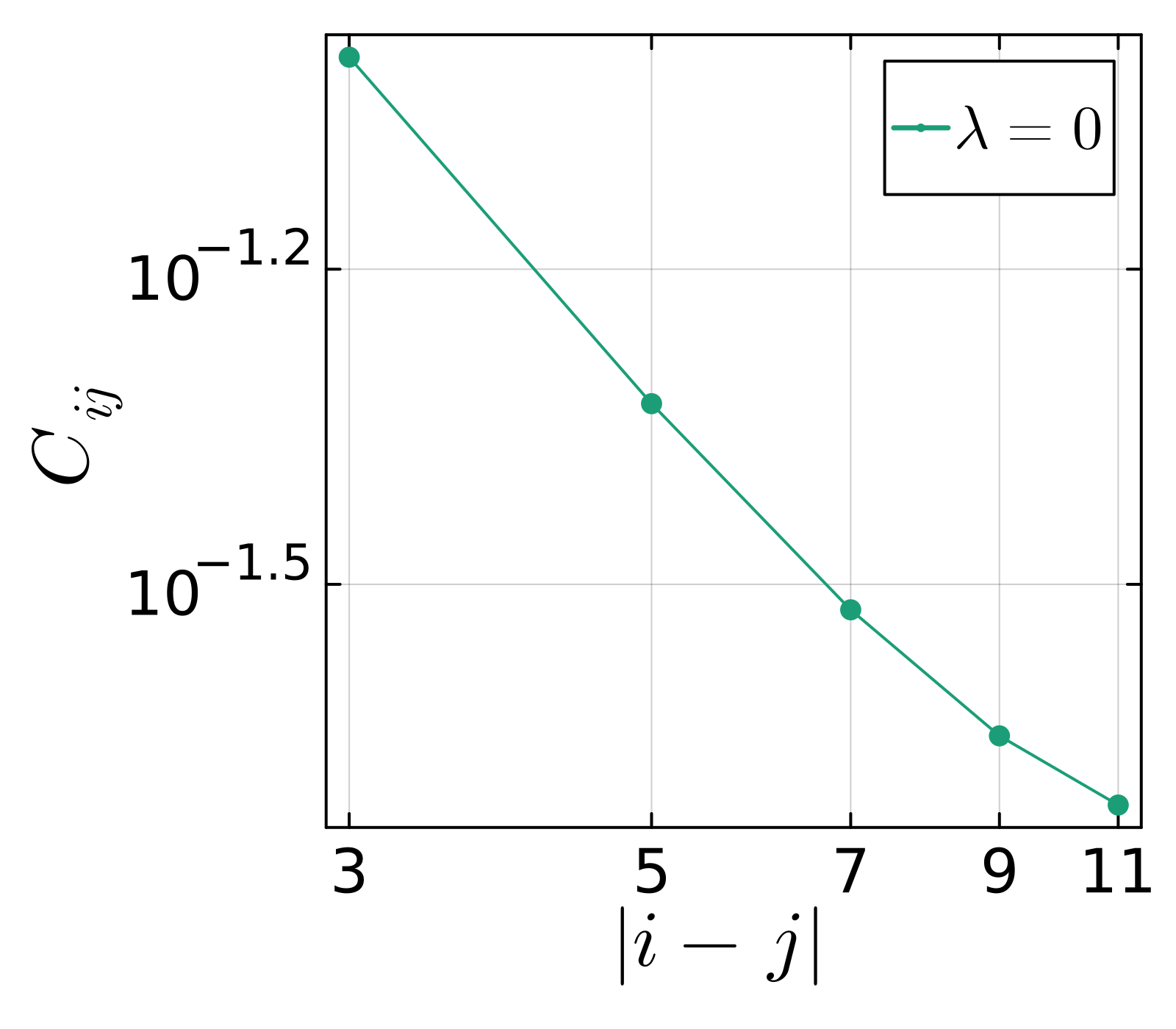}
    \includegraphics[width=0.49\linewidth]{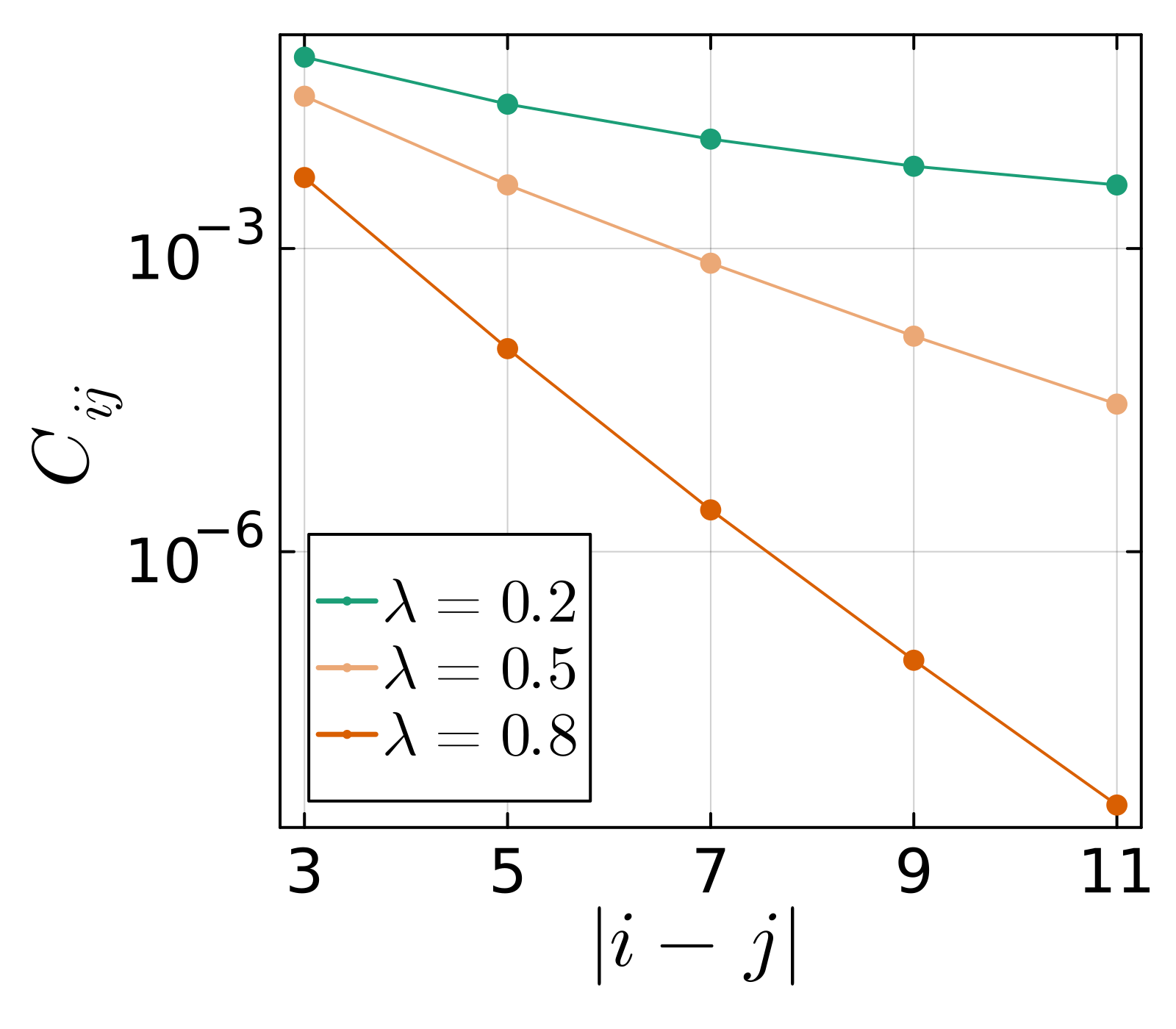}
    \caption{Two point correlation function $C_{ij}$ with respect to $|i-j|$, for the transition point $\lambda=0$ and non-transition points $\lambda=0.2,0.5,0.8$. The system size is $N=24$. The simulation is performed with ITensor Software Library~\cite{itensor} with the maximal bond dimension chosen as $D_{\mathrm{max}}=200$.}
    \label{fig:C}
\end{figure}

In our context, we are interested in the properties of $\rho_c$ as a density matrix, and whether $\rho_c$ has properties similar to those of $|\rho_c\drangle$ is unclear. As a first step investigation, we numerically computed the two-point correlation function of $\rho_c$,
\begin{equation}
    C_{ij} = \tr(\rho O_i O_j)-\tr(\rho O_i)\tr(\rho O_j).
\end{equation}
We present results at the critical point $\lambda=0$ and noncritical points $\lambda=0.2,0.5,0.8$ in Fig.~\ref{fig:C}, with $O$ chosen as the Hadamard gate. At $\lambda=0$, we observe that $C_{ij}$ exhibits power-law decay as a function of $|i-j|$. In contrast, away from criticality ($\lambda=0.2, 0.5, 0.8$), the correlations decay exponentially. Notably, the power-law decay at criticality forbids $\rho_c$ to be a Gibbs state of a local Hamiltonian. 

\section{Summary and outlook}
In this paper, we construct a quantum phase transition connecting two one-dimensional mixed-state renormalization fixed points, where one of which is intrinsically non-trivial and has a finite CMI. We provide a framework for rigorously establishing the phase equivalence by constructing the channel circuits. We demonstrate that the channel circuit constructed using the parent Lindbladian formalism preserves the correct symmetry and is crucial to justify the recovery channel circuits. 

There are many interesting open questions. In our model, we construct $\rho(\lambda)$ and prove its properties~\cref{eqn:rho-form,eqn:rho-property} starting from a two-dimensional PEPS tensor $A(\lambda)$ and the corresponding CP map $T(\lambda)$. A natural question to ask is whether one can construct a model where $\Lambda(\lambda)$ is given directly as a matrix product density operator with a finite bond dimension. Such a construction will make it possible to study the phase transition point analytically. We note that quantum phase transitions described by matrix product states with finite bond dimension have been studied in the literature~\cite{wolf2006quantum,jones2021skeletons}. It is natural to expect an analog to exist for matrix product density operators. One may also aim to construct a model that is not only continuous but also smooth in the parameter $\lambda$. 

Another interesting direction is to ask how to construct a phase transition between two fixed points $\rho_1$ and $\rho_2$ that are constructed from weak Hopf algebras. For example, when $\rho_1$ comes from the Fibonacci weak Hopf algebra. Our framework of proving phase equivalence relies on the fact that $\mathcal{E}$ and $\tE$ are finite-depth strictly-local channel circuits. This holds for $\rho_\ds$ whose bialgebraic structure $\mA$ is beyond weak Hopf algebra (discussed in~\cite{liu2025trading}) and the generalization to $\mathbb{Z}_\N$. For $\rho$ coming from a weak Hopf algebra (that is not a Hopf algebra), how to construct a finite-depth strictly-local channel circuit is not clear. One solution is to connect the algebra $\mA$ of $\rho$ with another one $\mA'$ that is beyond weak Hopf and construct the channel circuit via $\mA'$. We leave the framework of proving phase equivalence for this class of mixed states to future work. 

\section*{Ackowledgement}
We thank Ji-Yao Chen, Ignacio Cirac, Kohtaro Kato, Andras Molnar, Georgios Styliaris, and Xiehang Yu for insightful discussions. Y. L. is supported by the Alexander von Humboldt Foundation. 
\bibliography{ref}

\onecolumngrid
\appendix
  \section{Explicit construction of $\rho(\lambda)$}
\label{app:explicit}  
Consider the tensor network representation of the double semion ground state to toric code ground state phase transition on the square lattice~\cite{xu2018tensor}, whose local tensor $A(\lambda)$ is
\begin{equation}
\begin{aligned}
    \begin{array}{c}
        \begin{tikzpicture}[scale=1.,baseline={([yshift=-0.65ex] current bounding box.center)}]
       \pgfmathsetmacro{\rleg}{0.75}
        \pgfmathsetmacro{\ext}{0.35}
        \def\rc{0.25};
        \def\rdot{0.03}
    
        \opleg{(0,0)}{\rc}{}{\rleg};
    
        \coordinate (TR) at ({\rleg*cos(45)},  {\rleg*sin(45)});   
        \coordinate (TL) at ({\rleg*cos(135)}, {\rleg*sin(135)});  
        \coordinate (BR) at ({\rleg*cos(-45)}, {\rleg*sin(-45)});  
        \coordinate (BL) at ({\rleg*cos(225)}, {\rleg*sin(225)});  

        \filldraw[blue]  ({0, 0.6*\rc}) circle[radius=1.5*\rdot] node [above] {\scriptsize $I_2$};
         \filldraw[blue]  ({0.6*\rc,0}) circle[radius=1.5*\rdot] node [right] {\scriptsize $I_3$};
       \filldraw[blue]  ({0, -0.6*\rc}) circle[radius=1.5*\rdot] node [below] {\scriptsize $I_4$};
       \filldraw[blue]  ({-0.6*\rc,0}) circle[radius=1.5*\rdot] node [left] {\scriptsize $I_1$};

        \draw[Virtual] (TR) -- ++(\ext,0) node [right] {\scriptsize $\beta_1$};
        \draw[Virtual] (TR) -- ++(0,\ext) node [above] {\scriptsize $i_2$};
        \filldraw (TR) circle[radius=\rdot];

        \draw[Virtual] (TL) -- ++(-\ext,0) node [left] {\scriptsize $\alpha_1$};
        \draw[Virtual] (TL) -- ++(0,\ext) node [above] {\scriptsize $i_1$};
        \filldraw (TL) circle[radius=\rdot];

         \draw[Virtual] (BR) -- ++(\ext,0) node [right] {\scriptsize $\beta_2$};
        \draw[Virtual] (BR) -- ++(0,-\ext) node [below] {\scriptsize $i_2'$};
        \filldraw (BR) circle[radius=\rdot];

        \draw[Virtual] (BL) -- ++(-\ext,0) node [left] {\scriptsize $\alpha_2$};
        \draw[Virtual] (BL) -- ++(0,-\ext) node [below] {\scriptsize $i_1'$};
        \filldraw (BL) circle[radius=\rdot];

        \node[Virtual] [right] at ({0.6*\rleg*cos(45)},  {0.6*\rleg*sin(45)}) {\scriptsize $\beta$};
        \node[Virtual] [left] at ({0.6*\rleg*cos(135)},  {0.6*\rleg*sin(135)}) {\scriptsize $\alpha$};
        \node[Virtual] [left] at ({0.6*\rleg*cos(225)},  {0.6*\rleg*sin(225)}) {\scriptsize $\alpha'$};
        \node[Virtual] [right] at ({0.6*\rleg*cos(-45)},  {0.6*\rleg*sin(-45)}) {\scriptsize $\beta'$};
        
        \end{tikzpicture}
    \end{array}&=\sum_{\alpha\alpha'\beta\beta'}\Ainn_{\alpha\beta\alpha'\beta'}(\lambda) \delta_{ i_1 \alpha_1\alpha} \delta_{i_2 \beta_1 \beta} \delta_{i_1'\alpha_2  \alpha' }\delta_{i_2' \beta_2 \beta'}  D^{I_1}_{\alpha_1 \alpha_2} D^{I_2}_{i_1 i_2} D^{I_3}_{\beta_1 \beta_2} D^{I_4}_{i_1' i_2'}\\
    &:=A_{i_1 i_2 i_1' i_2'\alpha_1\alpha_2 \beta_1\beta_2}^{I_1 I_2 I_3 I_4} (\lambda).
\end{aligned}
\end{equation}
The blue dots $I_1,I_2,I_3,I_4$ represent physical degrees of freedom, and the red lines $i_1 i_2 i_1' i_2'\alpha_1\alpha_2 \beta_1\beta_2$ represent auxiliary degrees of freedom for the PEPS tensor. The black dots represent $\delta$ functions in the computational basis. 
The tensor $\Ainn$ only involves the auxiliary degrees of freedom, and the tensor $D$ involves both auxiliary degrees of freedom and the physical degree of freedom. The explicit form of $\Ainn(\lambda)$ is
\begin{equation}
\begin{aligned}
    \Ainn_{0011}&=\Ainn_{0101}=\lambda,\\
    \Ainn_{1100}&=\Ainn_{1010}=|\lambda|,\\
    \Ainn_{\alpha\beta\alpha'\beta'}&=1,\quad\text{otherwise},
\end{aligned}
\end{equation}
and the explicit form of $D$ is 
\begin{equation}
    \begin{aligned}
        D_{10}^1&=D_{01}^1 = D_{00}^0=D_{11}^0=1\\
        D_{ll'}^I&=0\quad\text{otherwise}. 
    \end{aligned}
\end{equation}
Physically, the tensor $D$ describes a domain wall when $I=1$. 
In this model, $\lambda=-1$ corresponds to the double semion phase, and $\lambda=1$ corresponds to the toric code phase. The phase transition point sits at $\lambda=0$. 

In the following, we will construct the transfer matrix $\tilde{E}(\lambda)$ using $A(\lambda)$, and connect to the tensor $E(\lambda)$ generating the matrix product CP map in the main text.

\subsection{Transfer matrix $\tilde{E}(\lambda)$}
Let's first consider the transfer matrix $E_0$ of the following tensor $A_0$
    \begin{equation}
\begin{aligned}
    \begin{array}{c}
        \begin{tikzpicture}[scale=1.,baseline={([yshift=-0.65ex] current bounding box.center)}]
       \pgfmathsetmacro{\rleg}{0.75}
        \pgfmathsetmacro{\ext}{0.35}
        \def\rc{0.25};
        \def\rdot{0.03}
    
        \opleg{(0,0)}{\rc}{}{\rleg};
    
        \coordinate (TR) at ({\rleg*cos(45)},  {\rleg*sin(45)});   
        \coordinate (TL) at ({\rleg*cos(135)}, {\rleg*sin(135)});  
        \coordinate (BR) at ({\rleg*cos(-45)}, {\rleg*sin(-45)});  
        \coordinate (BL) at ({\rleg*cos(225)}, {\rleg*sin(225)});  

        \filldraw[blue]  ({0, 0.6*\rc}) circle[radius=1.5*\rdot] node [above] {\scriptsize $I_2$};
         \filldraw[blue]  ({0.6*\rc,0}) circle[radius=1.5*\rdot] node [right] {\scriptsize $I_3$};
       \filldraw[blue]  ({0, -0.6*\rc}) circle[radius=1.5*\rdot] node [below] {\scriptsize $I_4$};
       \filldraw[blue]  ({-0.6*\rc,0}) circle[radius=1.5*\rdot] node [left] {\scriptsize $I_1$};

        \node[Virtual] [right] at ({0.6*\rleg*cos(45)},  {0.6*\rleg*sin(45)}) {\scriptsize $\beta$};
        \node[Virtual] [left] at ({0.6*\rleg*cos(135)},  {0.6*\rleg*sin(135)}) {\scriptsize $\alpha$};
        \node[Virtual] [left] at ({0.6*\rleg*cos(225)},  {0.6*\rleg*sin(225)}) {\scriptsize $\alpha'$};
        \node[Virtual] [right] at ({0.6*\rleg*cos(-45)},  {0.6*\rleg*sin(-45)}) {\scriptsize $\beta'$};
        
        \end{tikzpicture}
    \end{array}&=\Ainn_{\alpha\beta\alpha'\beta'}(\lambda)  D^{I_1}_{\alpha \alpha'} D^{I_2}_{\alpha\beta} D^{I_3}_{\beta \beta'} D^{I_4}_{\alpha'\beta'}=(A_0)_{\alpha\beta\alpha'\beta'}^{I_1 I_2 I_3 I_4} (\lambda).
\end{aligned}
\end{equation}
Define
\begin{equation}
    (E_0)_{\alpha\beta\alpha'\beta' \bar{\alpha}\bar{\beta}\bar{\alpha'}\bar{\beta'}}=\sum_{I_1 I_2 I_3 I_4}(A_0)_{\alpha\beta\alpha'\beta'}^{I_1 I_2 I_3 I_4}(\bar{A}_0)_{\bar{\alpha}\bar{\beta}\bar{\alpha'}\bar{\beta'}}^{I_1 I_2 I_3 I_4}
\end{equation}
where $\bar{A}_0$ is the complex conjugate of $A_0$. One can show
\begin{equation}
\begin{aligned}
      E_0&=(|0000\rangle+|1111\rangle)\otimes (\langle 0000|+\langle 1111|) + (|0001\rangle+|1110\rangle)\otimes(\langle 0001|+\langle 1110|)\\
      &\quad + (|0010\rangle+|1101\rangle)\otimes (\langle 0010|+\langle 1101|) + (|0100\rangle+|1011\rangle)\otimes(\langle 0100|+\langle 1011|)\\
      &\quad + (|1000\rangle+|0111\rangle)\otimes (\langle 1000|+\langle 0111|) +  (|0110\rangle+|1001\rangle)\otimes (\langle 0110|+\langle 1001|)\\
      &\quad + (\lambda |0011\rangle+|\lambda| |1100\rangle)\otimes (\lambda\langle 0011|+|\lambda|\langle 1100|)+(\lambda |0101\rangle+|\lambda| |1010\rangle)\otimes(\lambda \langle 0101|+|\lambda|\langle 1010|)\\
      &= \begin{cases}
          D_0 + X^{\otimes 4} D_0  & \lambda\geq 0\\
          D_0 + CZX^{(4)} D_0  & \lambda\leq 0
      \end{cases} 
\end{aligned}
\end{equation}
with 
\begin{equation}
    D_0(\lambda) := \bo-(1-\lambda^2)(|0011\rangle\langle 0011|+|1100\rangle\langle 1100|+|0101\rangle\langle0101|+|1010\rangle\langle 1010|),
\end{equation}
and $CZX^{(4)}:=(CZ)_{12}(CZ)_{23}(CZ)_{34}(CZ)_{41} X^{\otimes 4}$. Pictorially, 
\begin{equation}
\begin{aligned}
    &\left[D_0(\lambda)\right]_{\alpha\beta\alpha'\beta' \bar{\alpha}\bar{\beta}\bar{\alpha'}\bar{\beta'}}=\begin{array}{c}
        \begin{tikzpicture}[scale=1.,baseline={([yshift=-0.65ex] current bounding box.center)}]
         \pgfmathsetmacro{\rleg}{0.4}
        \pgfmathsetmacro{\rin}{0.7}
         \pgfmathsetmacro{\rout}{0.9}
        \pgfmathsetmacro{\mR}{0.8}
        \def\rdot{0.03}
            \opleg{(0,0)}{0.22}{$O_D$}{\rleg}

        \draw[\mthick, Virtual] ({\rleg*cos(135)},{\rleg*sin(135)}) .. controls ++(100:0.08) and ++(350:0.08) .. ({\rin*cos(135)},{\rin*sin(135)}) node [right] {\scriptsize $\alpha$};
        \draw[\mthick, Virtual,densely dashed] ({\rleg*cos(135)},{\rleg*sin(135)}) .. controls ++(170:0.2) and ++(270:0.2) .. ({\rout*cos(135)},{\rout*sin(135)}) node [left] {\scriptsize $\bar{\alpha}$};
        \filldraw ({\rleg*cos(135)},{\rleg*sin(135)}) circle[radius=\rdot];
    
        \draw[\mthick, Virtual] ({\rleg*cos(45)},{\rleg*sin(45)}) .. controls ++(80:0.08) and ++(190:0.08) .. ({\rin*cos(45)},{\rin*sin(45)}) node [left] {\scriptsize $\beta$};
        \draw[\mthick, Virtual,densely dashed] ({\rleg*cos(45)},{\rleg*sin(45)}) .. controls ++(10:0.2) and ++(270:0.2) .. ({\rout*cos(45)},{\rout*sin(45)}) node [right] {\scriptsize $\bar{\beta}$};
        \filldraw ({\rleg*cos(45)},{\rleg*sin(45)}) circle[radius=\rdot];
    
        \draw[\mthick, Virtual] ({\rleg*cos(225)},{\rleg*sin(225)}) .. controls ++(260:0.08) and ++(10:0.08) .. ({\rin*cos(225)},{\rin*sin(225)}) node [right] {\scriptsize $\alpha'$};
        \draw[\mthick, Virtual,densely dashed] ({\rleg*cos(225)},{\rleg*sin(225)}) .. controls ++(190:0.2) and ++(90:0.2) .. ({\rout*cos(225)},{\rout*sin(225)}) node [left] {\scriptsize $\bar{\alpha}'$};
        \filldraw ({\rleg*cos(225)},{\rleg*sin(225)}) circle[radius=\rdot];
    
        \draw[\mthick, Virtual] ({\rleg*cos(315)},{\rleg*sin(315)}) .. controls ++(280:0.08) and ++(170:0.08) .. ({\rin*cos(315)},{\rin*sin(315)}) node [left] {\scriptsize $\beta'$};
        \draw[\mthick, Virtual,densely dashed] ({\rleg*cos(315)},{\rleg*sin(315)}) .. controls ++(350:0.2) and ++(90:0.2) .. ({\rout*cos(315)},{\rout*sin(315)}) node [right] {\scriptsize $\bar{\beta}'$};
        \filldraw ({\rleg*cos(315)},{\rleg*sin(315)}) circle[radius=\rdot];
        \end{tikzpicture}
    \end{array},
\end{aligned}
\end{equation}
and
\begin{equation}
    X^{\otimes 4} D_0=\begin{array}{c}
        \begin{tikzpicture}[scale=1.,baseline={([yshift=-0.65ex] current bounding box.center)}]
         \pgfmathsetmacro{\rleg}{0.4}
        \pgfmathsetmacro{\rin}{0.8}
         \pgfmathsetmacro{\rout}{1.0}
        \pgfmathsetmacro{\mR}{0.8}
        \def\rdot{0.03}

        \opleg{(0,0)}{0.22}{$O_D$}{\rleg};
        \opEx{\rleg}{\rin}{\rout}{\mR}{\rdot};

        \op{({0.65*cos(45)-0.05},{0.65*sin(45)})}{0.06}{};
        \op{({0.65*cos(135)+0.05},{0.65*sin(135)})}{0.06}{};
        \op{({0.65*cos(225)+0.05},{0.65*sin(225)})}{0.06}{};
        \op{({0.65*cos(-45)-0.05},{0.65*sin(-45)})}{0.06}{};
        \end{tikzpicture}
    \end{array},\quad CZX^{(4)}D_0=\begin{array}{c}
        \begin{tikzpicture}[scale=1.,baseline={([yshift=-0.65ex] current bounding box.center)}]
         \pgfmathsetmacro{\rleg}{0.4}
        \pgfmathsetmacro{\rin}{0.95}
         \pgfmathsetmacro{\rout}{1.15}
        \pgfmathsetmacro{\mR}{0.8}
        \def\rdot{0.03}
            \opleg{(0,0)}{0.22}{$O_D$}{\rleg}
            \opEx{\rleg}{\rin}{\rout}{\mR}{\rdot};

        \draw[] ({0.65*cos(45)-0.05},{0.65*sin(45)}) -- ({0.65*cos(135)+0.05},{0.65*sin(135)});
        \draw[] ({0.65*cos(45)-0.05},{0.65*sin(45)}) -- ({0.65*cos(-45)-0.05},{0.65*sin(-45)});
        \draw[] ({0.65*cos(135)+0.05},{0.65*sin(135)}) -- ({0.65*cos(225)+0.05},{0.65*sin(225)});
        \draw[] ({0.65*cos(225)+0.05},{0.65*sin(225)}) -- ({0.65*cos(-45)-0.05},{0.65*sin(-45)});

        \opcolor{({0.65*cos(45)-0.05},{0.65*sin(45)})}{0.06}{}{lcolor};
        \opcolor{({0.65*cos(135)+0.05},{0.65*sin(135)})}{0.06}{}{lcolor};
        \opcolor{({0.65*cos(225)+0.05},{0.65*sin(225)})}{0.06}{}{lcolor};
        \opcolor{({0.65*cos(-45)-0.05},{0.65*sin(-45)})}{0.06}{}{lcolor};

        \op{({0.82*cos(45)-0.05},{0.82*sin(45)})}{0.06}{};
        \op{({0.82*cos(135)+0.05},{0.82*sin(135)})}{0.06}{};
        \op{({0.82*cos(225)+0.05},{0.82*sin(225)})}{0.06}{};
        \op{({0.82*cos(-45)-0.05},{0.82*sin(-45)})}{0.06}{};
        \end{tikzpicture}
    \end{array}
\end{equation}
Here the white circles denote the $X$ gates, and the two blue circles connected by a black line denotes a $CZ$ gate. 

The transfer matrix $\tilde{E}$ for tensor $A$ is defined in the same way,
\begin{equation}
    \left[\tilde{E}(\lambda)\right]^{i_1 i_2 i_1' i_2' j_1 j_2 j_1' j_2'}_{\alpha_1 \alpha_2 \beta_1\beta_2 \bar{\alpha}_1 \bar{\alpha}_2 \bar{\beta}_1  \bar{\beta}_2}=\sum_{I_1 I_2 I_3 I_4} A_{i_1 i_2 i_1' i_2'\alpha_1\alpha_2 \beta_1\beta_2}^{I_1 I_2 I_3 I_4} \bar{A}_{j_1 j_2 j_1' j_2'\bar{\alpha}_1\bar{\alpha}_2 \bar{\beta}_1\bar{\beta}_2}^{I_1 I_2 I_3 I_4}.
\end{equation}
With the transfer matrix $E_0$, the transfer matrix $\tilde{E}$ for tensor $A$ can be written easily by adding $\delta$ functions. Similar to $E_0$, $\tilde{E}$ consists of two pieces $\tilde{E}=\tilde{E}_1+\tilde{E}_2$. The first piece is
\begin{equation}
\begin{aligned}
    &\left[\tilde{E}_1(\lambda)\right]^{i_1 i_2 i_1' i_2' j_1 j_2 j_1' j_2'}_{\alpha_1 \alpha_2 \beta_1\beta_2 \bar{\alpha}_1 \bar{\alpha}_2 \bar{\beta}_1  \bar{\beta}_2}=\begin{array}{c}
        \begin{tikzpicture}[scale=1.,baseline={([yshift=-0.65ex] current bounding box.center)}]
         \pgfmathsetmacro{\rleg}{0.38}
        \pgfmathsetmacro{\rin}{0.6}
         \pgfmathsetmacro{\rout}{0.9}
        \pgfmathsetmacro{\mR}{1.1}
        \def\rdot{0.03}
            \opleg{(0,0)}{0.22}{$O_D$}{\rleg}
    \draw[\mthick,Virtual] ({\rin*cos(135)},{\rin*sin(135)})--({\mR*cos(180)},{\rin*sin(45)}) node [left] {\scriptsize $\alpha_1$};
    \draw[\mthick,Virtual] ({\rin*cos(135)},{\rin*sin(135)})--({-\rin*sin(45)},{\mR}) node [above] {\scriptsize $i_1$};
    \draw[\mthick,Virtual, dashed] ({\rout*cos(135)},{\rout*sin(135)}) -- ({\mR*cos(180)},{\rout*sin(45)}) node [left] {\scriptsize $\bar{\alpha}_1$};
    \draw[\mthick, Virtual, dashed] ({\rout*cos(135)},{\rout*sin(135)}) -- ({-\rout*sin(45)},{\mR}) node [above] {\scriptsize $j_1$};
    \draw[\mthick, Virtual] ({\rin*cos(135)},{\rin*sin(135)}) .. controls ++(350:0.08) and ++(100:0.08) .. ({\rleg*cos(135)},{\rleg*sin(135)});
    \draw[\mthick, Virtual, densely dashed] ({\rout*cos(135)},{\rout*sin(135)}) .. controls ++(250:0.2) and ++(170:0.2) .. ({\rleg*cos(135)},{\rleg*sin(135)});
     \filldraw ({\rin*cos(135)},{\rin*sin(135)}) circle[radius=\rdot];
     \filldraw ({\rout*cos(135)},{\rout*sin(135)}) circle[radius=\rdot];
     \filldraw ({\rleg*cos(135)},{\rleg*sin(135)}) circle[radius=\rdot];

    \draw[\mthick,Virtual] ({\rin*cos(45)},{\rin*sin(45)})--({\mR*cos(0)},{\rin*sin(45)}) node [right] {\scriptsize $\beta_1$};
    \draw[\mthick,Virtual] ({\rin*cos(45)},{\rin*sin(45)})--({\rin*sin(45)},{\mR}) node [above] {\scriptsize $i_2$};
    \draw[\mthick,Virtual, dashed] ({\rout*cos(45)},{\rout*sin(45)}) -- ({\mR*cos(0)},{\rout*sin(45)}) node [right] {\scriptsize $\bar{\beta}_1$};
    \draw[\mthick, Virtual, dashed] ({\rout*cos(45)},{\rout*sin(45)}) -- ({\rout*sin(45)},{\mR}) node [above] {\scriptsize $j_2$};
    \draw[\mthick, Virtual] ({\rin*cos(45)},{\rin*sin(45)}) .. controls ++(190:0.08) and ++(80:0.08) .. ({\rleg*cos(45)},{\rleg*sin(45)});
    \draw[\mthick, Virtual, densely dashed] ({\rout*cos(45)},{\rout*sin(45)}) .. controls ++(290:0.2) and ++(10:0.2) .. ({\rleg*cos(45)},{\rleg*sin(45)});
    \filldraw ({\rin*cos(45)},{\rin*sin(45)}) circle[radius=\rdot];
    \filldraw ({\rout*cos(45)},{\rout*sin(45)}) circle[radius=\rdot];
    \filldraw ({\rleg*cos(45)},{\rleg*sin(45)}) circle[radius=\rdot];

    \draw[\mthick,Virtual] ({\rin*cos(225)},{\rin*sin(225)})--({\mR*cos(180)},{-\rin*sin(45)}) node [left] {\scriptsize $\alpha_2$};
    \draw[\mthick,Virtual] ({\rin*cos(225)},{\rin*sin(225)})--({-\rin*sin(45)},{-\mR}) node [below] {\scriptsize $i_1'$};
    \draw[\mthick,Virtual, dashed] ({\rout*cos(225)},{\rout*sin(225)}) -- ({\mR*cos(180)},{-\rout*sin(45)}) node [left] {\scriptsize $\bar{\alpha}_2$};
    \draw[\mthick, Virtual, dashed] ({\rout*cos(225)},{\rout*sin(225)}) -- ({-\rout*sin(45)},{-\mR}) node [below] {\scriptsize $j_1'$};
    \draw[\mthick, Virtual] ({\rin*cos(225)},{\rin*sin(225)}) .. controls ++(10:0.08) and ++(260:0.08) .. ({\rleg*cos(225)},{\rleg*sin(225)});
    \draw[\mthick, Virtual, densely dashed] ({\rout*cos(225)},{\rout*sin(225)}) .. controls ++(70:0.2) and ++(190:0.2) .. ({\rleg*cos(225)},{\rleg*sin(225)});
    \filldraw ({\rin*cos(225)},{\rin*sin(225)}) circle[radius=\rdot];
    \filldraw ({\rout*cos(225)},{\rout*sin(225)}) circle[radius=\rdot];
    \filldraw ({\rleg*cos(225)},{\rleg*sin(225)}) circle[radius=\rdot];

    \draw[\mthick,Virtual] ({\rin*cos(315)},{\rin*sin(315)})--({\mR*cos(0)},{-\rin*sin(45)}) node [right] {\scriptsize $\beta_2$};
    \draw[\mthick,Virtual] ({\rin*cos(315)},{\rin*sin(315)})--({\rin*sin(45)},{-\mR}) node [below] {\scriptsize $i_2'$};
    \draw[\mthick,Virtual, dashed] ({\rout*cos(315)},{\rout*sin(315)}) -- ({\mR*cos(0)},{-\rout*sin(45)}) node [right] {\scriptsize $\bar{\beta}_2$};
    \draw[\mthick, Virtual, dashed] ({\rout*cos(315)},{\rout*sin(315)}) -- ({\rout*sin(45)},{-\mR}) node [below] {\scriptsize $j_2'$};
    \draw[\mthick, Virtual] ({\rin*cos(315)},{\rin*sin(315)}) .. controls ++(170:0.08) and ++(280:0.08) .. ({\rleg*cos(315)},{\rleg*sin(315)});
    \draw[\mthick, Virtual, densely dashed] ({\rout*cos(315)},{\rout*sin(315)}) .. controls ++(110:0.2) and ++(-10:0.2) .. ({\rleg*cos(315)},{\rleg*sin(315)});
    \filldraw ({\rin*cos(315)},{\rin*sin(315)}) circle[radius=\rdot];
    \filldraw ({\rout*cos(315)},{\rout*sin(315)}) circle[radius=\rdot];
    \filldraw ({\rleg*cos(315)},{\rleg*sin(315)}) circle[radius=\rdot];
    
        \end{tikzpicture}
    \end{array}=\delta_{i_1 j_1 \alpha_1\bar{\alpha}_1} \delta_{i_2 j_2 \beta_1 \bar{\beta}_1}\delta_{i_1' j_1' \alpha_2\bar{\alpha}_2 } \delta_{i_2' j_2' \beta_2 \bar{\beta}_2}(O_D)_{i_1 i_2 i_1' i_2'}, 
\end{aligned}
\end{equation}
with the black dots representing $\delta$ functions in the computational basis. In the region $\lambda\geq 0$, the second piece $\tilde{E}_2$ takes the form of
\begin{equation}
    \tilde{E}_2(\lambda)=\begin{array}{c}
        \begin{tikzpicture}[scale=1.,baseline={([yshift=-0.65ex] current bounding box.center)}]
         \pgfmathsetmacro{\rin}{0.6}
         \pgfmathsetmacro{\rout}{0.9}
        \pgfmathsetmacro{\mR}{1.1}
         \pgfmathsetmacro{\tes}{0.35}
        \opEtilde{(0,0)};
        \op{({\rin*cos(45)},{\rin*sin(45)+\tes})}{0.08}{};
         \op{({\rin*cos(135)},{\rin*sin(135)+\tes})}{0.08}{};
         \op{({\rin*cos(225)},{\rin*sin(225)-\tes})}{0.08}{};
         \op{({\rin*cos(-45)},{\rin*sin(-45)-\tes})}{0.08}{};

         \op{({\rin*cos(45)+\tes},{\rin*sin(45)})}{0.08}{};
         \op{({\rin*cos(135)-\tes},{\rin*sin(135)})}{0.08}{};
         \op{({\rin*cos(225)-\tes},{\rin*sin(225)})}{0.08}{};
         \op{({\rin*cos(-45)+\tes},{\rin*sin(-45)})}{0.08}{};

        \end{tikzpicture}
    \end{array}
\end{equation}
where the white circles denote the $X$ gates; and in the region $\lambda\leq 0$, the second piece is
\begin{equation}
    \tilde{E}_2(\lambda)=\begin{array}{c}
        \begin{tikzpicture}[scale=1.,baseline={([yshift=-0.65ex] current bounding box.center)}]
         \pgfmathsetmacro{\rin}{0.6}
         \pgfmathsetmacro{\rout}{0.9}
        \pgfmathsetmacro{\mR}{1.1}
         \pgfmathsetmacro{\tes}{0.5}
         \pgfmathsetmacro{\tesc}{0.3}
        \opEtilde{(0,0)};
        \op{({\rin*cos(45)},{\rin*sin(45)+\tes})}{0.08}{};
         \op{({\rin*cos(135)},{\rin*sin(135)+\tes})}{0.08}{};
         \op{({\rin*cos(225)},{\rin*sin(225)-\tes})}{0.08}{};
         \op{({\rin*cos(-45)},{\rin*sin(-45)-\tes})}{0.08}{};

         \op{({\rin*cos(45)+\tes},{\rin*sin(45)})}{0.08}{};
         \op{({\rin*cos(135)-\tes},{\rin*sin(135)})}{0.08}{};
         \op{({\rin*cos(225)-\tes},{\rin*sin(225)})}{0.08}{};
         \op{({\rin*cos(-45)+\tes},{\rin*sin(-45)})}{0.08}{};

        \coordinate (dotT1) at ({\rin*cos(45)},{\rin*sin(45)+\tesc});
        \coordinate (dotT2) at ({\rin*cos(135)},{\rin*sin(135)+\tesc});
        \draw[\mthick] (dotT1) -- (dotT2);
        \opcolor{dotT1}{0.08}{}{lcolor};
        \opcolor{dotT2}{0.08}{}{lcolor};
        
        \coordinate (dotB1) at ({\rin*cos(-45)},{\rin*sin(-45)-\tesc});
        \coordinate (dotB2) at ({\rin*cos(225)},{\rin*sin(225)-\tesc});
        \draw[\mthick] (dotB1) -- (dotB2);
        \opcolor{dotB1}{0.08}{}{lcolor};
        \opcolor{dotB2}{0.08}{}{lcolor};
        
        \coordinate (dotR1) at ({\rin*cos(45)+\tesc},{\rin*sin(45)});
        \coordinate (dotR2) at ({\rin*cos(-45)+\tesc},{\rin*sin(-45)});
        \draw[\mthick] (dotR1) -- (dotR2);
        \opcolor{dotR1}{0.08}{}{lcolor};
        \opcolor{dotR2}{0.08}{}{lcolor};
        
        \coordinate (dotL1) at ({\rin*cos(135)-\tesc},{\rin*sin(135)});
        \coordinate (dotL2) at ({\rin*cos(225)-\tesc},{\rin*sin(225)});
        \draw[\mthick] (dotL1) -- (dotL2);
        \opcolor{dotL1}{0.08}{}{lcolor};
        \opcolor{dotL2}{0.08}{}{lcolor};
        
        \end{tikzpicture}
    \end{array}.
\end{equation}
One can see that they coincide at $\lambda=0$ using the definition of $O_D$. 

Now, as in the main text, define the matrix product CP map $\tilde{T}(\lambda):=O^{(N)}(\tilde{E}(\lambda))$ as the horizontal concatenation of $N$ copies of $\tilde{E}$. From the structure of $\tilde{E}_1$ and $\tilde{E}_2$, one immediately see that
\begin{equation}
    \tilde{E}_2 \tilde{E}_1 =
    \begin{array}{c}
        \begin{tikzpicture}[scale=1.,baseline={([yshift=-0.65ex] current bounding box.center)}]
         \opEtilde{(0,0)};
         \pgfmathsetmacro{\rin}{0.6}
         \pgfmathsetmacro{\rout}{0.9}
        \pgfmathsetmacro{\mR}{1.1}
         \pgfmathsetmacro{\tes}{0.35}
        \opEtilde{(0,0)};
        \op{({\rin*cos(45)},{\rin*sin(45)+\tes})}{0.08}{};
         \op{({\rin*cos(135)},{\rin*sin(135)+\tes})}{0.08}{};
         \op{({\rin*cos(225)},{\rin*sin(225)-\tes})}{0.08}{};
         \op{({\rin*cos(-45)},{\rin*sin(-45)-\tes})}{0.08}{};

         \op{({\rin*cos(45)+\tes},{\rin*sin(45)})}{0.08}{};
         \op{({\rin*cos(135)-\tes},{\rin*sin(135)})}{0.08}{};
         \op{({\rin*cos(225)-\tes},{\rin*sin(225)})}{0.08}{};
         \op{({\rin*cos(-45)+\tes},{\rin*sin(-45)})}{0.08}{};
          \opEtilde{(2.2,0)};
        \end{tikzpicture}
    \end{array}
    =0
\end{equation}
due to the $X$ gates in $\tilde{E}_2$. Therefore, $\tilde{T}$ can be decomposed into $\tilde{T}=\tilde{T}_1+\tilde{T}_2$, where $\tilde{T}_a$ is generated by $\tilde{E}_a$ for $a=1,2$. Next, consider the concatenation of two $\tilde{E}_1$,
\[
\tilde{E}_1 \tilde{E}_1 =\begin{array}{c}
        \begin{tikzpicture}[scale=1.,baseline={([yshift=-0.65ex] current bounding box.center)}]
        \pgfmathsetmacro{\rin}{0.6}
         \pgfmathsetmacro{\rout}{0.9}
        \pgfmathsetmacro{\mR}{1.1}
         \opEtilde{(0,0)};
          \opEtilde{(2.2,0)};
          \node[Virtual] [above] at ({\rin*sin(45)-0.1},{\mR}) {\scriptsize $i_2$};
           \node[Virtual] [above] at ({\rout*sin(45)},{\mR}) {\scriptsize $j_2$};
           \node[Virtual] [above] at ({2.2+\rin*cos(135)+0.1},{\mR}) {\scriptsize $i_3$};
           \node[Virtual] [above] at ({2.2+\rout*cos(135)},{\mR}) {\scriptsize $j_3$};
        \end{tikzpicture}
    \end{array}.
\]
We see that the vertical degrees of freedom in the middle are redundant, in the sense that $i_2$ is enforced to be the same as $i_3$, and $j_2$ be the same as $j_3$ in the computational basis. We remove this redundancy by projecting $i_2,i_3$ to a single site, and projecting $j_2,j_3$ to a single site. Graphically, we replace the previous diagram by
\[
\tilde{E}_1 \tilde{E}_1\ra \begin{array}{c}
        \begin{tikzpicture}[scale=1.,baseline={([yshift=-0.65ex] current bounding box.center)}]
        \pgfmathsetmacro{\rleg}{0.38}
         \pgfmathsetmacro{\ext}{0.2}
        \pgfmathsetmacro{\x}{2*\rleg*cos(45)+3.0*\ext}
        \opEtilde[2]{(0,0)};
          \opEtilde[3]{(\x,0)};
\end{tikzpicture}
    \end{array}.
\]
By removing the redundant degrees of freedom for every site, we see that $E_1(\lambda)$ in~\cref{eqn:E1} generates the resulting CP map. Similarly, by considering the concatenation $\tilde{E}_2\tilde{E}_2$ and removing the redundancy, one obtain $E_2(\lambda)$ in the main text.

We comment that the row transfer matrix $\tilde{T}$, by construction, is a CP map that is a matrix product CP map. Therefore, $\tilde{\rho}_+$ is a fixed point of this CP map. Furthermore, one can bring $\tilde{T}$ to a quantum channel by a gauge transformation $\tilde{T}\ra (\sqrt{\Lambda}^{-1}\otimes \sqrt{\Lambda}^{-1})\tilde{T} (\sqrt{\Lambda}\otimes \sqrt{\Lambda})$, and if $\sqrt{\Lambda}$ allows an MPO representation, $\tilde{T}$ would be a matrix product quantum channel~\cite{stucchi2026structure}. 

\begin{figure}[t]
    \centering
    \includegraphics[width=0.395\linewidth]{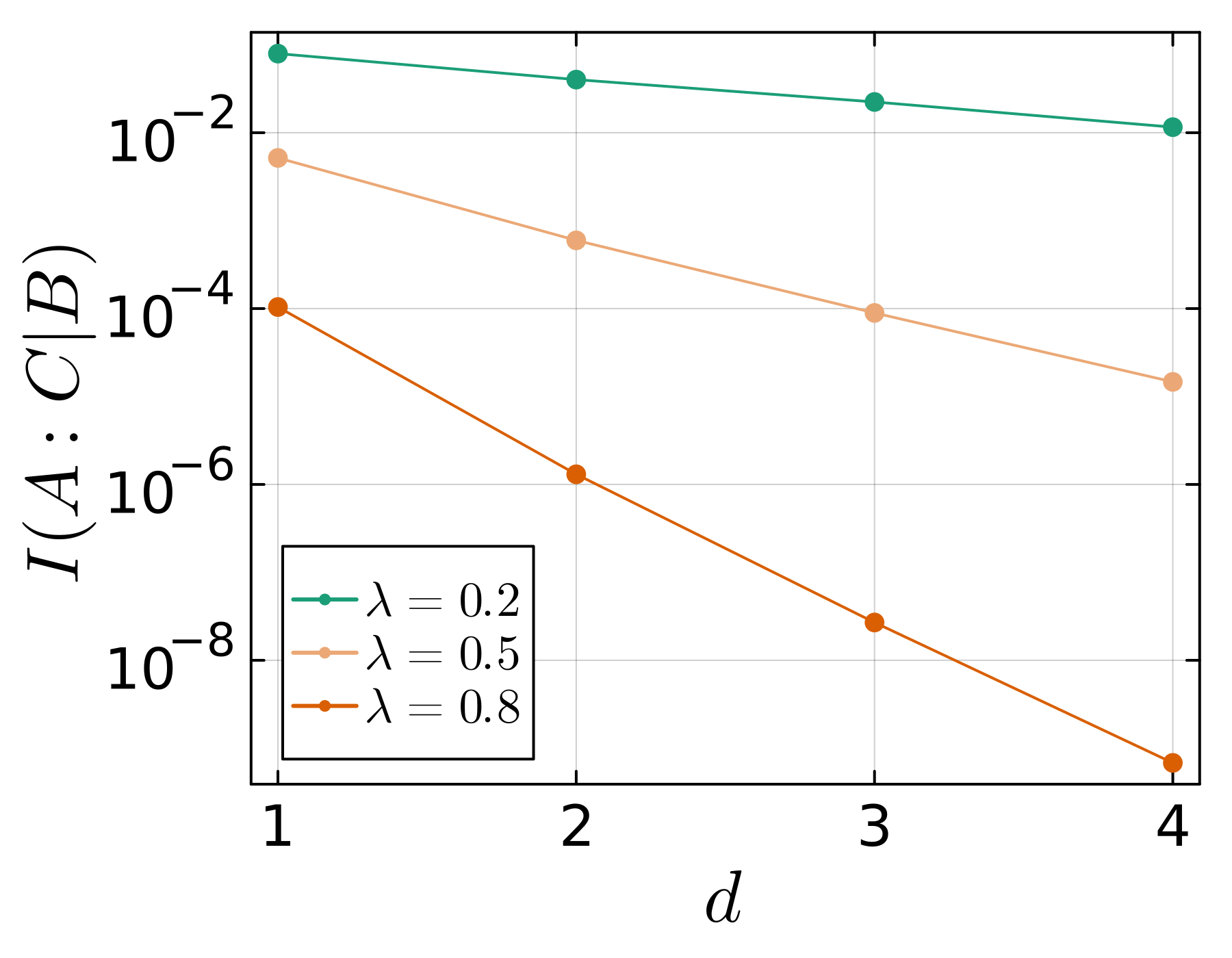}
    \includegraphics[width=0.4\linewidth]{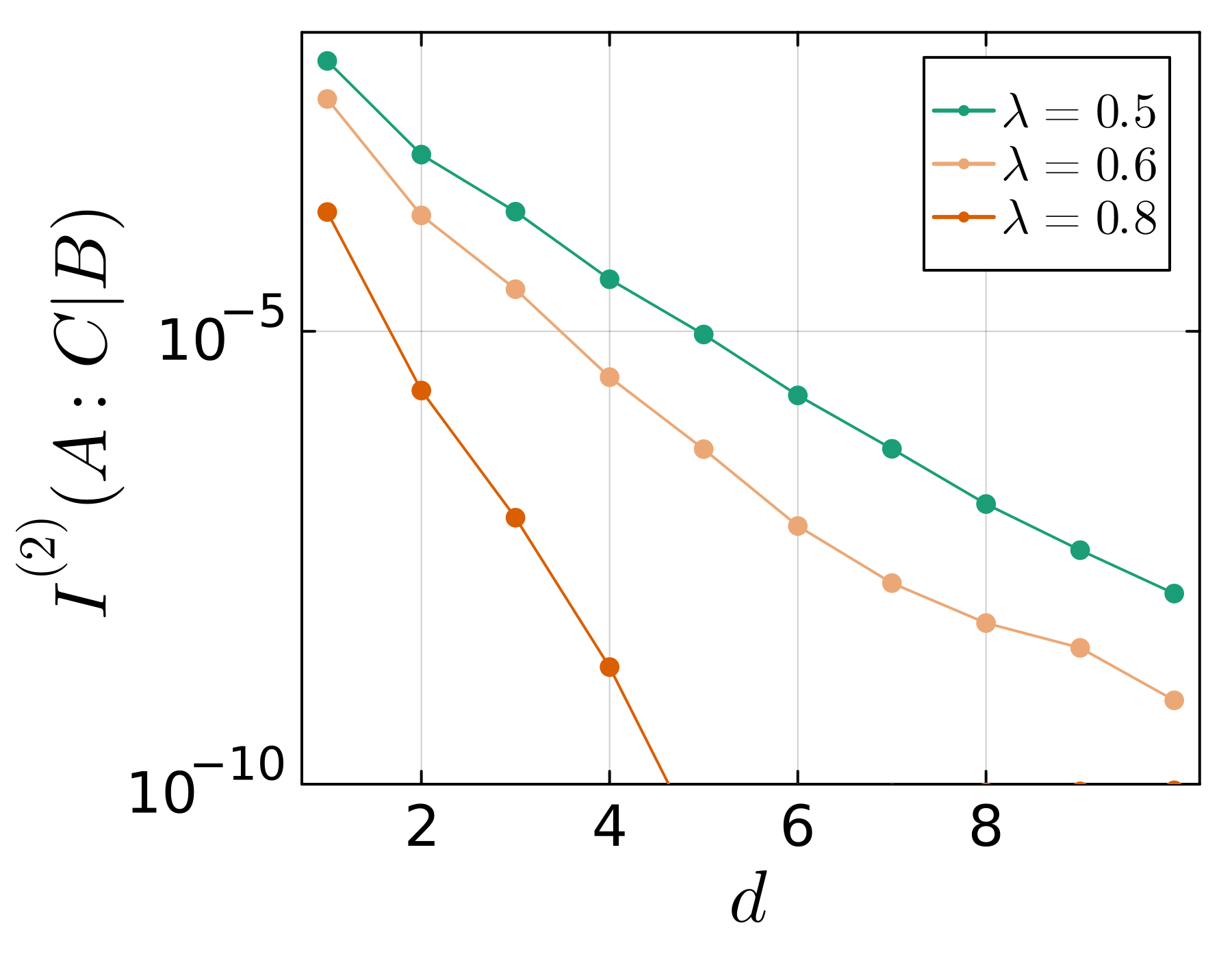}
    \caption{Left: Numerical result of conditional mutual information $I(A:C|B)$, for system size $N=12$ and $\lambda=0.2,0.5,0.8$. Right: Numerical result of Rényi-2 conditional mutual information $I^{(2)}(A:C|B)$, for system size $N=24$ and $\lambda=0.5,0.6,0.8$. We choose the size of region $A$ to be 2, and $d$ denotes the separation between region $A$ and $C$. The simulation is performed with ITensor Software Library with the maximal bond dimension chosen as $D_{\mathrm{max}}=200$.}
    \label{fig:CMI-numerics}
\end{figure}

\subsection{Decaying CMI of $\Lambda(\lambda)$}
\label{app:decay-CMI}

To justify that $\Lambda(\lambda)$ has a decaying CMI with a finite Markov length, we perform a numerical tensor-network simulation. We comment that it is not practical to expect that one can derive this analytically, by noting that $T_1^{(N)}$ is Hermitian, and one may interpret $-T_1^{(N)}$ as a Hamiltonian which allows an MPO representation. One may try to prove that this Hamiltonian has a gapped and unique ground state, and then conclude that this ground state has a finite correlation length. 
However, deciding whether an MPO Hamiltonian has a gapped unique ground state is a hard problem~\cite{bausch2020undecidability}. 

In the following, we provide numerical evidence for this model. For a mixed state, computing the conditional mutual information $I(A:C|B)$ is still difficult since it involves the computation of entanglement entropy, which requires exact diagonalization of the full reduced density matrix and is only feasible for small system size $N$. For a larger $N$, we evaluate the Rényi conditional mutual information. For Rényi index $\alpha$,
\begin{equation}
    I^{(\alpha)}(A:C|B):=S^{(\alpha)}(AB)+S^{(\alpha)}(BC)-S^{(\alpha)}(B)-S^{(\alpha)}(ABC)
\end{equation}
where $S^{(\alpha)}(\rho):=\frac{1}{1-\alpha}\log \tr(\rho^\alpha)$. We show the numerical result in Fig.~\ref{fig:CMI-numerics}. We observe that CMI for $N=12$ and Rényi-2 CMI for $N=24$ both decay with $d$, the size of region $B$ (separation between region $A$ and region $C$).



\section{Technical details of~\cref{sec:general}}
\label{app:details}
\subsection{Span of consecutive MPO tensors}
In this section, we prove~\cref{eqn:span-4-general}, the span of four consecutive MPO tensors, which relies on the structure of the underlying pre-bialgebra $\mA$~\cite{bohm_coassociativec_1996,bohm_weak_1999,bohm_weak_2000,molnar2022matrix}. 

Define the MPO generated by the tensor $A$ with boundary condition matrix $b$,
\begin{equation}
\begin{aligned}
     O^{(N)}(A;b)=&\sum_{\lbrace i,j\rbrace}\tr\left( A^{i_1 j_1} A^{i_2 j_2}\cdots A^{i_N j_N}b\right)\\
     &\quad  |i_1 i_2 \cdots i_N\rangle\langle j_1 j_2 \cdots j_N|,
\end{aligned}
\end{equation}
or graphically,
\begin{equation}
   O^{(N)}(A;b)=
    \begin{array}{c}
\begin{tikzpicture}[scale=1]
    \pgfmathsetmacro{\x}{0.494}       
    \pgfmathsetmacro{\dx}{0.988}      
    \pgfmathsetmacro{\xlast}{2.964}   
    \pgfmathsetmacro{\vl}{0.247}      
    \pgfmathsetmacro{\yo}{0.564}      
    \pgfmathsetmacro{\ybot}{0.318}    
    \pgfmathsetmacro{\ytop}{1.129}    
    \pgfmathsetmacro{\yleg}{0.811}    
    \pgfmathsetmacro{\ybez}{0.071}    

    \draw[Virtual] (\x+\vl, \yo)    -- (\x+\dx-\vl, \yo);   
    \draw[Virtual] (\x+\dx+\vl, \yo)-- (\x+\dx+\vl+\vl, \yo); 
    \node[anchor=center] at (\x+\dx+2*\vl+0.249, \yo+0.097) {$\overset{N}\cdots$};
    \draw[Virtual] (\xlast-\vl+\dx, \yo) -- (\xlast-\vl-\vl, \yo); 
    \draw[Virtual] (\x, \yo) -- (\x-\vl, \yo);               

    \draw[Virtual] (\x-\vl, \yo) arc[start angle=87.734, end angle=270, radius=\vl]
        -- (\xlast+\vl+\dx, \ybez) arc[start angle=-90, end angle=90, radius=\vl];

    \draw[-mid] (\x,      \yleg) -- (\x,      \ytop);
    \draw[-mid] (\x+\dx,  \yleg) -- (\x+\dx,  \ytop);
    \draw[-mid] (\xlast,  \yleg) -- (\xlast,  \ytop);

    \draw[Virtual] (\xlast+\vl, \yo) -- (\xlast, \yo);

    \draw[bevel, -mid] (\x,     0) -- (\x,     \ybot);
    \draw[bevel, -mid] (\x+\dx, 0) -- (\x+\dx, \ybot);
    \draw[bevel, -mid] (\xlast, 0) -- (\xlast, \ybot);

    \Osymb{(\x,     \yo)};
    \Osymb{(\x+\dx, \yo)};
    \Osymb{(\xlast, \yo)};
    \whaMsymborg{(\xlast+\dx, \yo)}{$b$}{white}
\end{tikzpicture}
        \end{array}
        \,.
\end{equation}

Let's first consider the span of two consecutive tensors $A^{(2)}_{g,\alpha\beta}$ (defined similarly as in~\cref{eqn:A4}, but with two sites), namely, the span of two consecutive sites with an arbitrary boundary condition. It was shown in~\cite{liu2025trading} that
\begin{equation}
    \mathrm{Span}\{A^{(2)}_{g,\alpha\beta} \}_{\alpha\beta}=\{|gk,gh\rangle\langle k,h| |k,h\in G\}. 
\end{equation}
Therefore, one can identify an algebra basis $\mB_g=\{e_{g,kh}\}_{kh}$, where each $e_{g,kh}\in\mA$ is associated with an independent boundary condition matrix $b(e_{g,kh})$, such that $O^{(2)}(A_g;b(e_{g,kh}))=|gk,gh\rangle\langle k,h|$. We denote 
\begin{equation}
    O^{(N)}[e_{g,kh}]:=O^{(N)}(A_g;b(e_{g,kh})).
\end{equation}

On this basis, the coproduct structure defined in~\cite{liu2025trading} reads
\begin{equation}
\Delta(e_{g,kl})=\sum_{mn}\frac{\omega(g,k,m-k)\omega(g,m,n-m)\omega(g,n,l-n)}{\omega(g,k,l-k)} e_{g,km}\otimes e_{g,nl}.
\end{equation}
Using that $O^{(4)}[e_{g,kl}]=O^{(2)}\otimes O^{(2)}[\Delta(e_{g,kl})]$, we read
\begin{equation}
\begin{aligned}
    O^{(4)}[e_{g,kl}]&=\sum_{mn}\frac{\omega(g,k,m-k)\omega(g,m,n-m)\omega(g,n,l-n)}{\omega(g,k,l-k)}|gk,gm,gn,gl\rangle\langle k,m,n,l|\\
    &=(X^g)^{\otimes 4} D_g^{(4)} (|k\rangle\langle k|\otimes \bo\otimes \bo\otimes |l\rangle\langle l|). 
\end{aligned}
\end{equation}
Identify 
\begin{equation}
     \mathrm{Span}\{A^{(4)}_{g,\alpha\beta} \}_{\alpha\beta}=\{O^{(4)}[e_{g,kl}]\}_{kl}
\end{equation}
leads to~\cref{eqn:span-4-general}. 

\subsection{Proof of~\cref{eqn:commute-general}}
In this section, we prove
\begin{equation}
    [D_{g_1}^{(4)},(X^{g_2})^{\otimes 4}]=0,\quad \forall g_1, g_2
\end{equation}
Since $D_{g_1}^{(4)}$ is diagonal in the computational basis with 
\begin{equation}
    D_{g_1}^{(4)}|k,m,n,l\rangle=d_{g_1}(k,m,n,l)|k,m,n,l\rangle
\end{equation}
and $(X^{g_2})^{\otimes 4}|k,m,n,l\rangle=|k+g_2,m+g_2,n+g_2,l+g_2\rangle$, the commutation condition is equivalent to
\begin{equation}
    d_{g_1}(k+g_2,m+g_2,n+g_2,l+g_2)=d_{g_1}(k,m,n,l). 
\end{equation}

Note that for $k,m\in\{0,\cdots,n-1\}$, we have $[k]_n+[m-k]_n-[[k]_n+[m-k]_n]_n=k+[m-k]_n-m$, therefore
\begin{equation}
\begin{aligned}
    \ln d_{g_1}(k,m,n,l)&=\frac{2\pi i g_1}{n^2}\left(
    (k+[m-k]_n-m)+(m+[n-m]_n-n)+(n+[l-n]_n -l)-(k+[l-k]_n-l)
    \right)\\
    &=\frac{2\pi i g_1}{n^2}([m-k]_n+[n-m]_n + [l-n]_n-[l-k]_n)
\end{aligned}
\end{equation}
which only depends on the differences between two arguments. Therefore $d_{g_1}(k+g_2,m+g_2,n+g_2,l+g_2)=d_{g_1}(k,m,n,l)$. 

\end{document}

%% file: my_tikz.tex
\newcommand\mthick{}

\newcommand{\op}[3]{
\begin{scope}[shift={(#1)}]
	\filldraw[\mthick, fill=white](0,0) circle[radius=#2];
	\draw (0,0) node {\scriptsize #3};
\end{scope}
}

\newcommand{\opcolor}[4]{
\begin{scope}[shift={(#1)}]
	\filldraw[\mthick, fill=#4](0,0) circle[radius=#2];
	\draw (0,0) node {\scriptsize #3};
\end{scope}
}

\newcommand{\opleg}[4]{
\begin{scope}[shift={(#1)}]
	\draw[\mthick, Virtual] ({#2*cos(45)},{#2*sin(45)}) -- ({#4*cos(45)},{#4*sin(45)});
	\draw[\mthick, Virtual] ({#2*cos(135)},{#2*sin(135)}) -- ({#4*cos(135)},{#4*sin(135)});
	\draw[\mthick, Virtual] ({#2*cos(225)},{#2*sin(225)}) -- ({#4*cos(225)},{#4*sin(225)});
	\draw[\mthick, Virtual] ({#2*cos(315)},{#2*sin(315)}) -- ({#4*cos(315)},{#4*sin(315)});
	\filldraw[\mthick, fill=white](0,0) circle[radius=#2];
	\draw (0,0) node {\scriptsize #3};
\end{scope}
}

\newcommand{\opE}[1]{
\begin{scope}[shift={(#1)}]
    \pgfmathsetmacro{\rleg}{0.35}
    \pgfmathsetmacro{\ext}{0.2}
    \def\rdot{0.03}

    \opleg{(0,0)}{0.22}{$O_D$}{\rleg};

    \coordinate (TR) at ({\rleg*cos(45)},  {\rleg*sin(45)});   
    \coordinate (TL) at ({\rleg*cos(135)}, {\rleg*sin(135)});  
    \coordinate (BR) at ({\rleg*cos(-45)}, {\rleg*sin(-45)});  
    \coordinate (BL) at ({\rleg*cos(225)}, {\rleg*sin(225)});  

   \coordinate (JTR) at ({\rleg*cos(45)+1.5*\ext},  {\rleg*sin(45)+\ext*0.707});
    \coordinate (JTL) at ({\rleg*cos(135)-1.5*\ext}, {\rleg*sin(135)+\ext*0.707});
    \coordinate (JBR) at ({\rleg*cos(-45)+1.5*\ext}, {\rleg*sin(-45)-\ext*0.707});
    \coordinate (JBL) at ({\rleg*cos(225)-1.5*\ext}, {\rleg*sin(225)-\ext*0.707});

    \draw[\mthick, Virtual] (TR) to[out=45,  in=180] (JTR);
    \draw[\mthick, dashed]          (JTR) to[out=45,  in=270] ++({ \ext*0.707}, {2.*\ext}) ; 
    \draw[\mthick]  (JTR) to[out=135, in=270] ++({-\ext*0.707}, {2.*\ext}) ; 
    \filldraw (JTR) circle[radius=\rdot];

    \draw[\mthick, Virtual] (TL) to[out=135, in=0]   (JTL);

    \draw[\mthick, Virtual] (BR) to[out=-45, in=180] (JBR);
    \draw[\mthick, dashed]          (JBR) to[out=-45, in=90] ++({ \ext*0.707},{-2.*\ext}); 
    \draw[\mthick]  (JBR) to[out=-135, in=90] ++({-\ext*0.707},{-2.*\ext}); 
    \filldraw (JBR) circle[radius=\rdot];

    \draw[\mthick, Virtual] (BL) to[out=225, in=0]   (JBL) ;
\end{scope}
}

\newcommand{\opEtilde}[2][1]{
\begin{scope}[shift={#2}]
 \pgfmathsetmacro{\rleg}{0.38}
 \pgfmathsetmacro{\ext}{0.2}
        \pgfmathsetmacro{\rin}{0.6}
         \pgfmathsetmacro{\rout}{0.9}
        \pgfmathsetmacro{\mR}{1.1}
        \def\rdot{0.03}
            \opleg{(0,0)}{0.22}{$O_D$}{\rleg}

    \coordinate (TR) at ({\rleg*cos(45)},  {\rleg*sin(45)});   
    \coordinate (TL) at ({\rleg*cos(135)}, {\rleg*sin(135)});  
    \coordinate (BR) at ({\rleg*cos(-45)}, {\rleg*sin(-45)});  
    \coordinate (BL) at ({\rleg*cos(225)}, {\rleg*sin(225)});  

   \coordinate (JTR) at ({\rleg*cos(45)+1.5*\ext},  {\rleg*sin(45)+\ext*0.707});
    \coordinate (JTL) at ({\rleg*cos(135)-1.5*\ext}, {\rleg*sin(135)+\ext*0.707});
    \coordinate (JBR) at ({\rleg*cos(-45)+1.5*\ext}, {\rleg*sin(-45)-\ext*0.707});
    \coordinate (JBL) at ({\rleg*cos(225)-1.5*\ext}, {\rleg*sin(225)-\ext*0.707});
            
     \ifnum#1=3
        \draw[\mthick, Virtual] (TL) to[out=135, in=0]   (JTL);
     \else
        \draw[\mthick,Virtual] ({\rin*cos(135)},{\rin*sin(135)})--({\mR*cos(180)},{\rin*sin(45)});
        \draw[\mthick,Virtual] ({\rin*cos(135)},{\rin*sin(135)})--({-\rin*sin(45)},{\mR});
        \draw[\mthick,Virtual, dashed] ({\rout*cos(135)},{\rout*sin(135)}) -- ({\mR*cos(180)},{\rout*sin(45)});
        \draw[\mthick, Virtual, dashed] ({\rout*cos(135)},{\rout*sin(135)}) -- ({-\rout*sin(45)},{\mR});
        \draw[\mthick, Virtual] ({\rin*cos(135)},{\rin*sin(135)}) .. controls ++(350:0.08) and ++(100:0.08) .. ({\rleg*cos(135)},{\rleg*sin(135)});
        \draw[\mthick, Virtual, densely dashed] ({\rout*cos(135)},{\rout*sin(135)}) .. controls ++(250:0.2) and ++(170:0.2) .. ({\rleg*cos(135)},{\rleg*sin(135)});
         \filldraw ({\rin*cos(135)},{\rin*sin(135)}) circle[radius=\rdot];
         \filldraw ({\rout*cos(135)},{\rout*sin(135)}) circle[radius=\rdot];
         \filldraw ({\rleg*cos(135)},{\rleg*sin(135)}) circle[radius=\rdot];
    \fi

    \ifnum#1=2
        \draw[\mthick, Virtual] (TR) to[out=45,  in=180] (JTR);
        \draw[\mthick, dashed, Virtual]          (JTR) to[out=45,  in=270] ++({ \ext*0.707}, {3.*\ext}) ; 
        \draw[\mthick, Virtual]  (JTR) to[out=135, in=270] ++({-\ext*0.707}, {3.*\ext}) ; 
        \filldraw (JTR) circle[radius=\rdot];
    \else
        \draw[\mthick,Virtual] ({\rin*cos(45)},{\rin*sin(45)})--({\mR*cos(0)},{\rin*sin(45)});
        \draw[\mthick,Virtual] ({\rin*cos(45)},{\rin*sin(45)})--({\rin*sin(45)},{\mR});
        \draw[\mthick,Virtual, dashed] ({\rout*cos(45)},{\rout*sin(45)}) -- ({\mR*cos(0)},{\rout*sin(45)});
        \draw[\mthick, Virtual, dashed] ({\rout*cos(45)},{\rout*sin(45)}) -- ({\rout*sin(45)},{\mR});
        \draw[\mthick, Virtual] ({\rin*cos(45)},{\rin*sin(45)}) .. controls ++(190:0.08) and ++(80:0.08) .. (TR);
        \draw[\mthick, Virtual, densely dashed] ({\rout*cos(45)},{\rout*sin(45)}) .. controls ++(290:0.2) and ++(10:0.2) .. (TR);
        \filldraw ({\rin*cos(45)},{\rin*sin(45)}) circle[radius=\rdot];
        \filldraw ({\rout*cos(45)},{\rout*sin(45)}) circle[radius=\rdot];
        \filldraw ({\rleg*cos(45)},{\rleg*sin(45)}) circle[radius=\rdot];
    \fi

    \ifnum#1=3
        \draw[\mthick, Virtual] (BL) to[out=225, in=0]   (JBL) ;
    \else
        \draw[\mthick,Virtual] ({\rin*cos(225)},{\rin*sin(225)})--({\mR*cos(180)},{-\rin*sin(45)}) ;
        \draw[\mthick,Virtual] ({\rin*cos(225)},{\rin*sin(225)})--({-\rin*sin(45)},{-\mR});
        \draw[\mthick,Virtual, dashed] ({\rout*cos(225)},{\rout*sin(225)}) -- ({\mR*cos(180)},{-\rout*sin(45)});
        \draw[\mthick, Virtual, dashed] ({\rout*cos(225)},{\rout*sin(225)}) -- ({-\rout*sin(45)},{-\mR}) ;
        \draw[\mthick, Virtual] ({\rin*cos(225)},{\rin*sin(225)}) .. controls ++(10:0.08) and ++(260:0.08) .. (BL);
        \draw[\mthick, Virtual, densely dashed] ({\rout*cos(225)},{\rout*sin(225)}) .. controls ++(70:0.2) and ++(190:0.2) .. (BL);
        \filldraw ({\rin*cos(225)},{\rin*sin(225)}) circle[radius=\rdot];
        \filldraw ({\rout*cos(225)},{\rout*sin(225)}) circle[radius=\rdot];
        \filldraw ({\rleg*cos(225)},{\rleg*sin(225)}) circle[radius=\rdot];
    \fi

    \ifnum#1=2
        \draw[\mthick, Virtual] (BR) to[out=-45, in=180] (JBR);
        \draw[\mthick, dashed, Virtual]          (JBR) to[out=-45, in=90] ++({ \ext*0.707},{-3.*\ext}); 
        \draw[\mthick, Virtual]  (JBR) to[out=-135, in=90] ++({-\ext*0.707},{-3.*\ext}); 
        \filldraw (JBR) circle[radius=\rdot];
    \else
        \draw[\mthick,Virtual] ({\rin*cos(315)},{\rin*sin(315)})--({\mR*cos(0)},{-\rin*sin(45)}) ;
        \draw[\mthick,Virtual] ({\rin*cos(315)},{\rin*sin(315)})--({\rin*sin(45)},{-\mR}) ;
        \draw[\mthick,Virtual, dashed] ({\rout*cos(315)},{\rout*sin(315)}) -- ({\mR*cos(0)},{-\rout*sin(45)}) ;
        \draw[\mthick, Virtual, dashed] ({\rout*cos(315)},{\rout*sin(315)}) -- ({\rout*sin(45)},{-\mR});
        \draw[\mthick, Virtual] ({\rin*cos(315)},{\rin*sin(315)}) .. controls ++(170:0.08) and ++(280:0.08) .. (BR);
        \draw[\mthick, Virtual, densely dashed] ({\rout*cos(315)},{\rout*sin(315)}) .. controls ++(110:0.2) and ++(-10:0.2) .. (BR);
        \filldraw ({\rin*cos(315)},{\rin*sin(315)}) circle[radius=\rdot];
        \filldraw ({\rout*cos(315)},{\rout*sin(315)}) circle[radius=\rdot];
        \filldraw ({\rleg*cos(315)},{\rleg*sin(315)}) circle[radius=\rdot];
    \fi
\end{scope}
}

\newcommand{\opEx}[5]{
    \draw[\mthick, Virtual] ({#1*cos(135)},{#1*sin(135)}) .. controls ++(100:0.08) and ++(350:0.08) .. ({#2*cos(135)},{#2*sin(135)}) ;
    \draw[\mthick, Virtual,densely dashed] ({#1*cos(135)},{#1*sin(135)}) .. controls ++(170:0.2) and ++(270:0.2) .. ({#3*cos(135)},{#3*sin(135)});
    \filldraw ({#1*cos(135)},{#1*sin(135)}) circle[radius=#5];
    \draw[\mthick, Virtual] ({#1*cos(45)},{#1*sin(45)}) .. controls ++(80:0.08) and ++(190:0.08) .. ({#2*cos(45)},{#2*sin(45)});
    \draw[\mthick, Virtual,densely dashed] ({#1*cos(45)},{#1*sin(45)}) .. controls ++(10:0.2) and ++(270:0.2) .. ({#3*cos(45)},{#3*sin(45)});
    \filldraw ({#1*cos(45)},{#1*sin(45)}) circle[radius=#5];
    \draw[\mthick, Virtual] ({#1*cos(225)},{#1*sin(225)}) .. controls ++(260:0.08) and ++(10:0.08) .. ({#2*cos(225)},{#2*sin(225)});
    \draw[\mthick, Virtual,densely dashed] ({#1*cos(225)},{#1*sin(225)}) .. controls ++(190:0.2) and ++(90:0.2) .. ({#3*cos(225)},{#3*sin(225)});
    \filldraw ({#1*cos(225)},{#1*sin(225)}) circle[radius=#5];
    \draw[\mthick, Virtual] ({#1*cos(315)},{#1*sin(315)}) .. controls ++(280:0.08) and ++(170:0.08) .. ({#2*cos(315)},{#2*sin(315)});
    \draw[\mthick, Virtual,densely dashed] ({#1*cos(315)},{#1*sin(315)}) .. controls ++(350:0.2) and ++(90:0.2) .. ({#3*cos(315)},{#3*sin(315)});
    \filldraw ({#1*cos(315)},{#1*sin(315)}) circle[radius=#5];
}

\newcommand{\opsc}[5]{
\begin{scope}[shift={(#1)}]
	\draw[ \mthick, fill=#5, rounded corners=1pt] (-#2,-#3) rectangle (#2,#3);
	\draw (0,0) node {\scriptsize #4};
\end{scope}
}

\newcommand{\TCZket}[2]{
\begin{scope}[shift={(#1)}]
\def\disp{0.3};
        \def\rd{0.15};
        \def\rt{3.0};
        \def\rtt{5.5};
        \def\off{0.12};
        \def\linethick{};

            \draw[Virtual, \linethick, rounded corners=2pt]
              ({-\disp},{\disp-\off/2}) -- ({-\rt*\disp},{\rt*\disp-\off/2}) -- ({-\rt*\disp},{\rtt*\disp});
            \draw[Virtual, \linethick, rounded corners=2pt]
               ({\disp},{\disp+\off/2}) -- ({\rt*\disp-\off},{\rt*\disp+\off/2}) -- ({\rt*\disp-\off},{\rtt*\disp});
            \draw[Virtual, \linethick, rounded corners=2pt]
              ({-\disp},{-\disp+\off/2}) -- ({-\rt*\disp},{-\rt*\disp+\off/2}) -- ({-\rt*\disp},{-\rtt*\disp});
            \draw[Virtual, \linethick, rounded corners=2pt]
            ({\disp},{-\disp-\off/2}) -- ({\rt*\disp-\off},{-\rt*\disp-\off/2}) -- ({\rt*\disp-\off},{-\rtt*\disp});

            \op{(-\disp,\disp)}{\rd}{\scriptsize $X$};
            \op{(\disp,\disp)}{\rd}{\scriptsize $X$};
            \op{(-\disp,-\disp)}{\rd}{\scriptsize $X$};
            \op{(\disp,-\disp)}{\rd}{\scriptsize $X$};

            \opsc{(-\off/2,2.1*\disp)}{0.65}{0.12}{}{lcolor};
            \opsc{(-\off/2,-2.1*\disp)}{0.65}{0.12}{}{lcolor};
            \ifnum#2=1
            \opsc{(2.1*\disp,0)}{0.12}{0.6}{}{lcolor};
            \opsc{(-2.1*\disp,0)}{0.12}{0.6}{}{lcolor};
            \fi
            \ifnum#2=2
            \opsc{(2.1*\disp,0)}{0.12}{0.6}{}{lcolor};
            \fi
            \ifnum#2=3
            \opsc{(-2.1*\disp,0)}{0.12}{0.6}{}{lcolor};
            \fi
\end{scope}
}

\newcommand{\GcTensor}[6]{
    \begin{scope}[shift={(#1)}]
    \ifnum#5=0
		\draw[Virtual] (-#2,0) -- (#2,0);
		\draw[] (0,#2) -- (0,-#2);
    \fi
    \ifnum#5=-1
		\draw[Virtual] (0,0) -- (#2,0);
		\draw[] (0,#2) -- (0,-#2);
    \fi
    \ifnum#5=1
		\draw[Virtual] (-#2,0) -- (0,0);
		\draw[] (0,#2) -- (0,-#2);
    \fi

    \ifnum#5=2
		\draw[Virtual] (-#2,0) -- (#2,0);
		\draw[] (0,0) -- (0,#2);
    \fi
    \ifnum#5=-2
		\draw[Virtual] (-#2,0) -- (#2,0);
		\draw[] (0,0) -- (0,-#2);
    \fi

    \ifnum#5=3
		\draw[Virtual] (-#2,0) -- (#2,0);
		\draw[] (0,-#2) -- (0,#2);
    \fi
    \ifnum#5=4
    \fi
        \draw[fill=#6, rounded corners=2pt,\mthick] (-#3,-#3) rectangle (#3,#3);
		\draw (0,0) node {\scriptsize #4};
	\end{scope}
}

\newcommand{\whaMsymborg}[3]{
\begin{scope}[shift={(#1)}]
\filldraw[fill=#3] 
    (0,0) circle[radius=0.247];
    \node[anchor=center,scale=0.9] at (0,0) {#2};
\end{scope}
}

\newcommand{\Osymb}[1]{
\whaMsymborg{#1}{$A$}{white}
}

\newcommand{\whaMorg}[4]{
\begin{scope}[shift={(#1)}]
\def\rd{0.247};
\def\rlen{0.318};
\ifnum#2=1
    \draw[-mid] (0, \rd) -- (0, \rd+\rlen);
    \draw[-mid] (0, -\rd-\rlen) -- (0, -\rd);
    \draw[Virtual, -mid] (\rd+\rlen, 0) -- (\rd, 0);
    \draw[Virtual, -mid] (-\rd, 0) -- (-\rd-\rlen, 0);
\fi
\ifnum#2=2
    \draw[-mid] (0, \rd) -- (0, \rd+\rlen);
    \draw[-mid] (0, -\rd-\rlen) -- (0, -\rd);
    \draw[Virtual] (\rd+\rlen, 0) -- (\rd, 0);
    \draw[Virtual] (-\rd, 0) -- (-\rd-\rlen, 0);
\fi
\whaMsymborg{(0,0)}{#3}{#4};
\end{scope}
}

\newcommand{\mposite}[2]{
\whaMorg{#1}{#2}{$A$}{white};
}

\newcommand{\whaATensor}[3]{
\begin{scope}[shift={(#1)}]
\def\rd{0.247};
\def\rlen{0.318};

\ifnum#2=1
\draw[] (0, \rd) -- (0, \rd+\rlen);
\draw[Virtual] (\rd+\rlen, 0) -- (\rd, 0);
\draw[Virtual] (-\rd, 0) -- (-\rd-\rlen, 0);
\fi
\ifnum#2=3
\draw[] (0, -\rd) -- (0, -\rd-\rlen);
\draw[Virtual] (\rd+\rlen, 0) -- (\rd, 0);
\draw[Virtual] (-\rd, 0) -- (-\rd-\rlen, 0);
\fi
\draw[fill=tensorcolor, rounded corners=2pt,\mthick] (-\rd,-\rd) rectangle (\rd,\rd);
\node[anchor=center] at (0,0) {#3};

\end{scope}
}